\documentclass{jfm}
\pdfoutput=1
\usepackage{graphicx}
\usepackage{xcolor}
\usepackage{natbib}
\usepackage{amsmath}
\usepackage{bm}

\newcommand{\ie}{i.e.,\ }
\newcommand{\eg}{e.g.,\ }

\newcommand{\figrefC}[2][]{\mbox{Figure \ref{#2}(#1)}}
\newcommand{\figrefAC}[1]{\mbox{Figure \ref{#1}}}

\newcommand{\equref}[1]{\mbox{equation (\ref{#1})}}
\newcommand{\figref}[2][]{\mbox{figure \ref{#2}(#1)}}
\newcommand{\figrefA}[1]{\mbox{figure \ref{#1}}}
\newcommand{\tabref}[1]{\mbox{table \ref{#1}}}
\newcommand{\secref}[1]{\mbox{section \ref{#1}}}

\title[Turbulent channel flow over a viscous hyper-elastic wall]
{Numerical simulation of turbulent channel flow over a viscous hyper-elastic wall}

\author[M.E. Rosti \& L. Brandt ]{Marco E. Rosti\thanks{Email address for correspondence: merosti@kth.se} and Luca Brandt}

\affiliation{Linn\'{e} Flow Centre and SeRC (Swedish e-Science Research Centre), \\KTH Mechanics, SE 100 44 Stockholm, Sweden}

\begin{document}

\maketitle

\begin{abstract}
We perform numerical simulations of a turbulent channel flow over an hyper-elastic wall. In the fluid region the flow is governed by the incompressible Navier--Stokes (NS) equations, while the solid is a neo-Hookean material satisfying the incompressible Mooney-Rivlin law. The multiphase flow is solved with a one-continuum formulation, using a monolithic velocity field for both the fluid and solid phase, which allows the use of a fully Eulerian formulation. The simulations are carried out at Reynolds bulk $Re=2800$ and examine the effect of different elasticity and viscosity of the deformable wall. We show that the skin friction increases monotonically with the material elastic modulus. The turbulent flow in the channel is affected by the moving wall even at low values of elasticity since non-zero fluctuations of vertical velocity at the interface influence the flow dynamics. The near-wall streaks and the associated quasi-streamwise vortices are strongly reduced near a highly elastic wall while the flow becomes more correlated in the spanwise direction, similarly to what happens for flows over rough and porous walls. As a consequence, the mean velocity profile in wall units is shifted downwards when shown in logarithmic scale, and the slope of the inertial range increases in comparison to that for the flow over a rigid wall. We propose a correlation between the downward shift of the inertial range, its slope and the wall-normal velocity fluctuations at the wall, extending results for the flow over rough walls. We finally show that the interface deformation is determined by the fluid fluctuations when the viscosity of the elastic layer is low, while when this is high the deformation is limited by the solid properties.
\end{abstract}

\section{Introduction} \label{sec:introduction}
\subsection{Aim and objectives}
Near-wall turbulence is responsible for significant drag penalties in many flows of engineering relevance. Because of that, many researchers have studied the flow over complex walls as found in many industrial and natural flows, e.g.\ porous, rough and deformable to mention few examples \citep{luo_bewley_2005a, breugem_boersma_uittenbogaard_2006a, orlandi_leonardi_2008a, pluvinage_kourta_bottaro_2014a, rosti_cortelezzi_quadrio_2015a}. These investigations aim  to understand the effect of such complex boundaries on the flow field and inspire the design of novel materials able to modify the flow and reduce drag. The ability to design materials with specific properties (\eg deformability, porosity, permeability, \ldots) could lead to novel developments in several fields spanning from aerodynamics to biology, from chemistry to medicine. Among the many control strategies, the use of compliant surfaces is one of the most attractive since no energy in introduced in the system and the structure is allowed to deform in response to the fluctuations of the near-wall turbulence. In this context, the aim of this work is to explore and better understand the interactions between the turbulent flow and a deformable wall.

Materials for which the constitutive behaviour is only a function of the current state of deformation are generally known as elastic. In the special case when the work done by the stresses during a deformation process is dependent only on the initial and final configurations, the behaviour of the material is path-independent and a stored strain energy function or elastic potential can be defined \citep{bonet_wood_1997a}. These so-called hyper-elastic materials show non-linear stress-strain curves and are generally used to describe rubber-like substances; these constitutive laws are employed in this work.

\subsection{Flow over rigid complex walls}
Many applications involve fluid flows over or through porous and rough materials. In his pioneering work on turbulence over rough walls, \citet{nikuradse_1933a, nikuradse_1950a}  presented a large number of experimental measurements in pipes with walls covered by sand grain. He identified  three regimes by plotting the friction factor versus the Reynolds number, $Re$: in the first one, at low $Re$, the friction follows the law of laminar smooth walls, and does not depend on the roughness. In the transitional regime, the friction depends on $Re$ and on the kind of roughness, and finally, at higher $Re$, the friction depends only on the kind of roughness, and not on the Reynolds number, a regime defined as fully rough. More recently, \citet{cabal_szumbarski_floryan_2002a} considered the instability of the flow over wavy surfaces and found that a two-dimensional disturbance leads to the formation of three-dimensional vortical structures near the wall that are the precursor of the streaky structures observed in turbulent smooth-wall flows. \citet{leonardi_orlandi_smalley_djenidi_antonia_2003a} studied the fully turbulent flow over rough walls by means of direct numerical simulations of a channel flow with spanwise-aligned square bars on one of the walls. These authors were able to quantify the importance of the form drag, relative to the skin-frictional drag, from the pressure distribution around the bars, and were also able to assess how low- and high-speed streaks are disrupted by the roughness elements \citep{leonardi_orlandi_djenidi_antonia_2004a}. Contour plots of the normal component of vorticity in the near-wall region indicated that the turbulent structures become shorter and wider than over a smooth wall, \ie the wall layer tends towards isotropy, an observation consistent with the previous laboratory observations of \citet{antonia_krogstad_2001a}.

In an attempt to model the mean flow properties, \citet{clauser_1954a} showed that the effect of the roughness could be quantified by a shift $\Delta U^+$ of the mean velocity distribution in the logarithmic region \citep{hama_1954a}; this shift is assumed to depend on the density and shape of the roughness elements and on the Reynolds number. Moreover, the effective origin of the mean velocity profile is taken at a distance $d$ from the roughness crest plane, which can be defined in several ways. \citet{orlandi_leonardi_antonia_2006a} performed DNS of turbulent channel flows with square, circular and triangular rods to explore the possibility of an universal parametrization with quantities related to the flow within the roughness, motivated by the common belief that geometrical parameters are not sufficient for classifying the large variety of roughness \citep{belcher_jerram_hunt_2003a}. Indeed, they observed a poor correlation between $\Delta U^+$ and the geometrical parameters, while a satisfactory collapse was achieved by plotting $\Delta U^+$ versus the rms of the normal velocity component at the plane of the roughness crests.  \citet{orlandi_leonardi_tuzi_antonia_2003a} demonstrated that the normal velocity distribution on the plane of the crests is the driving mechanism for the modifications of the  near-wall structures.

As concerns the flow over a porous material, the main effects are the destabilization of laminar flows and the enhancement of the Reynolds-shear stresses, with a consequent increase in skin-friction drag in turbulent flows. The first results showing the destabilizing effects of the wall permeability were obtained experimentally by \citet{beavers_sparrow_magnuson_1970a}, whereas more recently \citet{tilton_cortelezzi_2006a, tilton_cortelezzi_2008a} performed a three-dimensional temporal linear stability analysis of a laminar flow in a channel with one or two homogeneous, isotropic, porous walls. These authors show that wall permeability can drastically decrease the stability of fully developed laminar channel flows. Numerical and experimental works have also been performed to understand the effects of permeability on turbulent flows. \citet{breugem_boersma_uittenbogaard_2006a} studied the influence of a highly permeable porous wall, made by a packed bed of particles, on turbulent channel flows. Their results showed that the structure and dynamics of the turbulence above a highly permeable wall are different from those of a turbulent flow over an effectively impermeable wall due to the strong reduction of the intensity of the low- and high-speed streaks and of the quasi-streamwise vortices characteristics of wall-bounded flows. Moreover, relatively large spanwise vortical structures are found to increase the exchange of momentum between the porous medium and the channel, thus inducing a strong increase in the Reynolds-shear stresses and, consequently, in the skin friction. \citet{rosti_cortelezzi_quadrio_2015a} extended the analysis to a porous material with relatively small permeability, where inertial effects can be neglected in the porous material, and  separately examined the effect of porosity and permeability. \citet{suga_matsumura_ashitaka_tominaga_kaneda_2010a} studied experimentally the effects of wall permeability on a turbulent flow in a channel with a porous wall and observed that the slip velocity over a permeable wall increases drastically in the range of Reynolds numbers where the flow transitions from laminar to turbulent. In addition, the transition to turbulence appears at progressively lower Reynolds numbers as permeability increases, consistently with the results of linear stability analysis \citep{tilton_cortelezzi_2006a, tilton_cortelezzi_2008a}. The turbulence statistics of the velocity fluctuations showed that the wall-normal component increases as the wall permeability and/or the Reynolds number increases. \citet{suga_matsumura_ashitaka_tominaga_kaneda_2010a} performed a numerical simulation of the same turbulent flow using an analytic wall function at the interface and found the results in good agreement with their experimental data and the numerical simulations by \citet{breugem_boersma_uittenbogaard_2006a}.

\subsection{Flow over deformable complaint walls}
As discussed above, rough and porous walls have a destabilising effect on the flow, and the near-wall low- and high-speed streaks have significantly lower amplitude than over an impermeable smooth wall. Also, the wall-normal component of the velocity at the interface is found to be strongly correlated to the flow behaviour. We will now discuss the flow over a deformable wall, when the wall-normal component of the velocity is not zero because of the movement of the wall.

The flow over a deformable moving wall differs from that past rigid surfaces, because of the two-way coupling between the flow and wall dynamics. In particular, it has been shown that the elasticity of the flexible surface alters the transition from laminar to turbulent flow. The first experimental results by \citet{lahav_eliezer_silberberg_1973a} and the following study by \citet{krindel_silberberg_1979a} showed that flows through gel-coated tubes are subject to a considerable drag increase, suggesting that this is due to a transition to the turbulent flow at  low Reynolds numbers ($Re \approx 600$). Later on, several researchers studied the linear stability of a fluid flow through flexible channels and pipes with elastic and hyper-elastic walls \citep{kumaran_fredrickson_pincus_1994a, srivatsan_kumaran_1997a, shankar_kumaran_1999a, kumaran_1995a, kumaran_1996a, kumaran_1998a, kumaran_1998b, kumaran_muralikrishnan_2000a}, in the limit of small disturbances superimposed to the laminar flow, considering linearized governing equations. The main result of these studies is the possibility of the flow to be unstable even in the absence of fluid inertia, \ie at zero Reynolds number. In particular, \citet{kumaran_fredrickson_pincus_1994a} and \citet{kumaran_1995a} studied the linear stability of Couette and pipe flow over a visco-elastic medium in the zero Reynolds number limit, reporting that the flow can be unstable also in these geometries. The authors explain that the instability arises due to the energy transfer from the mean flow to the fluctuations because of the deformation work at the interface. Different instabilities have been found in this kind of flows and been classified as follows: $i)$ viscous instabilities, $ii)$ low Reynolds number long-wave instabilities, $iii)$ rigid surface modes, $iv)$ regular inviscid modes, $v)$ singular inviscid modes, and $vi)$ high Reynolds number wall modes. The interested reader is referred to \cite{shankar_kumaran_1999a} for more details. Recently, \citet{verma_kumaran_2013a} found a reduction in the transition Reynolds number and fast mixing in a micro-channel due to a dynamical instability induced by a soft wall. The soft wall is made of soft gel, and good agreement is found with their results from linear stability and numerical simulation where the gel is modelled as an incompressible viscous hyper-elastic material.

Nonetheless, scarce is the work done on the effect of elastic surfaces on the turbulent flow. \citet{luo_bewley_2003a, luo_bewley_2005a} considered a new class of compliant surfaces, called tensegrity fabrics, formed as a stable pre-tensioned network of compressive members (bars) interconnected by tensile members (tendons), when no individual structural member ever experiences bending moments. Their simulations showed that the interface forms a streamwise travelling wave convected at a high phase velocity for low stiffness and damping of the structure. By performing a parametric study of the material parameters, these authors found a resonating condition between the wall deformation and the turbulent flow; the wavy motion of the interface caused large increases of the drag and of the turbulent kinetic energy of the flow. Notwithstanding that compliant surfaces can delay the transition to turbulence \citep{benjamin_1960a, landahl_1962a, carpenter_garrad_1985a, carpenter_morris_1990a, davies_carpenter_1997a, daniel_gaster_willis_1987a, gaster_1988a}, most experiments have not been able to find a reduction in the turbulence-induced drag \citep{bushnell_hefner_ash_1977a, carpenter_garrad_1985a, gad-el-hak_1986a, gad-el-hak_1987a, gad-el-hak_others_1996a}, with a couple of exceptions, such as \citet{lee_fisher_schwarz_1993a} and \citet{choi_yang_clayton_glover_atlar_semenov_kulik_1997a} who observed a reduction of turbulent intensity and drag in their experiments of boundary layers over compliant surfaces.

\subsection{Outline}
In this work, we present the first Direct Numerical Simulations (DNS) of turbulent channel flow at a bulk Reynolds number of $Re=2800$ bounded by an incompressible hyper elastic wall. In the fluid part of the channel the full incompressible Navier-Stokes equations are solved, while momentum conservation and the incompressibility constraint are enforced inside the solid material.  In \secref{sec:formulation}, we first discuss the flow configuration and governing equations, and then present the numerical methodology used. A validation of the numerical implementation is reported in \secref{sec:validation}, while the effects of an hyper-elastic wall on a fully developed turbulent channel flow are presented  in \secref{sec:result}, where  we also discuss the role of the different parameters defining the elastic wall. Finally, a summary of the main findings and some conclusions are drawn  in \secref{sec:conclusion}.

\begin{figure}
  \centering
  \includegraphics[width=0.4\textwidth]{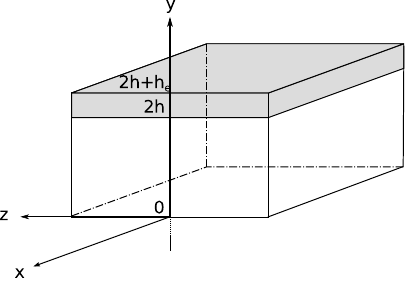}
  \caption{Sketch of the channel geometry. The interface between the channel region and the elastic medium is located at $y=2h$, while two solid rigid walls bounds the domain at $y=0$ and $2h+h_e$.}
  \label{fig:sketch}
\end{figure}

\section{Formulation} \label{sec:formulation}
We consider the turbulent flow of an incompressible viscous fluid through a channel with an incompressible hyper elastic wall. \figrefAC{fig:sketch} shows a sketch of the geometry and the Cartesian coordinate system, where $x$, $y$ and $z$ ($x_1$, $x_2$, and $x_3$) denote the streamwise, wall-normal and spanwise coordinates, while $u$, $v$ and $w$ ($u_1$, $u_2$, and $u_3$) denote the respective components of the velocity vector field. The upper interface is initially located at $y=2h$, while the lower and upper impermeable walls are located at $y=0$ and $2h+h_e$, respectively. Hence, $h_e$ represents the height of the hyper-elastic layer.

The fluid and solid phase motion is governed by conservation of momentum and the incompressibility constraint:
\begin{subequations}
\label{eq:NS}
\begin{align}
\frac{\partial u_i^f}{\partial t} + \frac{\partial u_i^f u_j^f}{\partial x_j} &= \frac{1}{\rho} \frac{\partial \sigma_{ij}^f}{\partial x_j}, \\
\frac{\partial u_i^f}{\partial x_i} &= 0, \\
\frac{\partial u_i^s}{\partial t} + \frac{\partial u_i^s u_j^s}{\partial x_j} &= \frac{1}{\rho} \frac{\partial \sigma_{ij}^s}{\partial x_j}, \\
\frac{\partial u_i^s}{\partial x_i} &= 0,
\end{align}
\end{subequations}
where the suffixes $^f$ and $^s$ are used to indicate the fluid and solid phase. In the previous set of equations, $\rho$ is the density (assumed to be the same for the solid and fluid), and $\sigma_{ij}$ the Cauchy stress tensor. The kinematic and dynamic interactions between the fluid and solid phases are determined by enforcing the continuity of the velocity and traction force at the interface between the two phases
\begin{subequations}
\label{eq:bc}
\begin{align}
u_i^f &= u_i^s, \label{bc-v}\\
\sigma_{ij}^f n_j &= \sigma_{ij}^s n_j \label{bc-sigma},
\end{align}
\end{subequations}
where $n_i$ denotes the normal vector at the interface.

To numerically solve the fluid-structure interaction problem at hand, we use the so called one-continuum formulation \citep{tryggvason_sussman_hussaini_2007a}, where only one set of equations is solved over the whole domain. This is achieved by introducing a monolithic velocity vector field $u_i$ valid everywhere; this is found by applying the volume averaging procedure \citep{takeuchi_yuki_ueyama_kajishima_2010a, quintard_whitaker_1994b} and reads
\begin{equation}
\label{eq:phi-v}
u_i = \left( 1 - \phi^s \right) u_i^f + \phi^s u_i^s,
\end{equation}
where $\phi^s$ is the solid volume fraction. This is zero in the fluid, $\phi^s=0$, whereas $\phi^s=1$ in the solid, with $0\le\phi^s\le1$ close to the interface. In particular, the isoline at $\phi^s=1/2$ represents the interface. Similarly to the Volume of Fluid \citep{hirt_nichols_1981a} and Level Set \citep{sussman_smereka_osher_1994a, chang_hou_merriman_osher_1996a} methods used to simulate multi phase flows, we can write the stress in a mixture form as
\begin{equation}
\label{eq:phi-stress}
\sigma_{ij} = \left( 1 - \phi^s \right) \sigma_{ij}^f + \phi^s \sigma_{ij}^s.
\end{equation}

The fluid is assumed to be Newtonian so that the stress in the fluid 
\begin{equation}
\label{eq:stress-f}
\sigma_{ij}^f = -p \delta_{ij} + 2 \mu^f D_{ij},
\end{equation}
where $p$ is the pressure, $\mu^f$ the fluid dynamic viscosity, $D_{ij}$ the strain rate tensor defined as
\begin{equation}
\label{eq:strainrate}
D_{ij}=\frac{1}{2} \left( \frac{\partial u_i}{\partial x_j} + \frac{\partial u_j}{\partial x_i} \right),
\end{equation}
and $\delta$ is the Kronecker delta. The solid is an incompressible viscous hyper-elastic material undergoing only the isochoric motion with constitutive equation
\begin{equation}
\label{eq:stress-s}
\sigma_{ij}^s = -p \delta_{ij} + 2 \mu^s D_{ij} + \sigma_{ij}^{sh},
\end{equation}
where $\mu^s$ is the solid dynamic viscosity, and the last term the hyper-elastic contribution modelled as a neo-Hookean material, thus satisfying the incompressible Mooney-Rivlin law. The constitutive equation for the solid becomes
\begin{equation}
\label{eq:stress-s-he}
\sigma_{ij}^s = -p \delta_{ij} + 2 \mu^s D_{ij} + G B_{ij}
\end{equation}
where $B_{ij}$ is the left Cauchy-Green deformation tensor, and $G$ the modulus of transverse elasticity. The solid constitutive equation is a function of the left Cauchy-Green deformation tensor, and the set of equations for the solid material can be closed in a purely Eulerian manner by updating its component with the following transport equation:
\begin{equation}
\label{eq:B-adv}
\frac{\partial B_{ij}}{\partial t} + \frac{\partial u_k B_{ij}}{\partial x_k} = B_{kj}\frac{\partial u_i}{\partial x_k} + B_{ik}\frac{\partial u_j}{\partial x_k}.
\end{equation}
The previous equation comes from the fact that the upper convected derivative of the left Cauchy-Green deformation tensor is identically zero \citep{bonet_wood_1997a}. To close the full system, one transport equation is needed for the solid volume fraction $\phi^s$,
\begin{equation}
\label{eq:PHI-adv}
\frac{\partial \phi^s}{\partial t} + \frac{\partial u_k \phi^s}{\partial x_k} = 0.
\end{equation}

\begin{figure}
  \centering
  \includegraphics[width=0.5\textwidth]{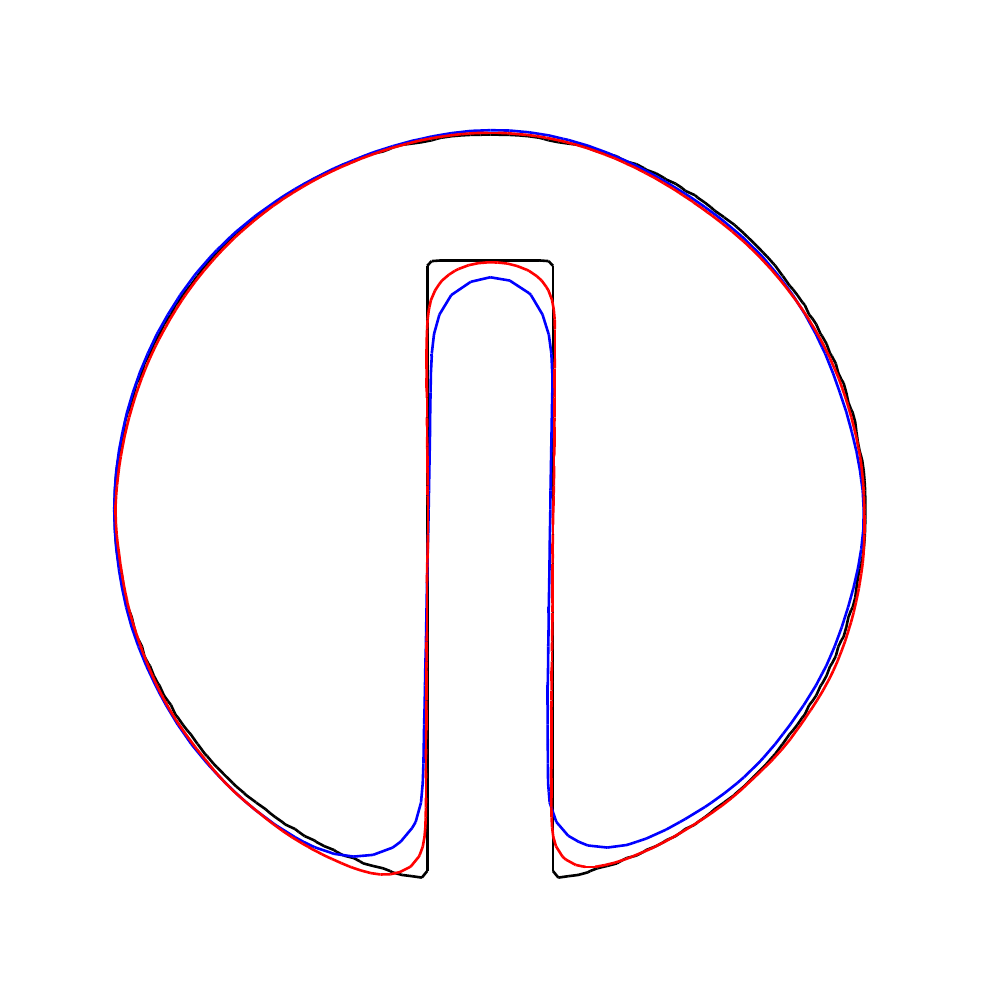}
  \caption{The Zalesak's disk: validation of the current implementation by simulating a  slotted disc undergoing solid body rotation, see \cite{zalesak_1979a}. The black line denotes the exact solution, whereas blue and red the solution obtained with $64$ and $128$ grid points per diameter.}
  \label{fig:zalesak}
\end{figure}

\subsection{Numerical implementation}
In order to simulate the elastic material, we follow the method for fluid-structure interaction problems recently developed by \citet{sugiyama_ii_takeuchi_takagi_matsumoto_2011a}. These authors propose  a fully Eulerian formulation on a fixed Cartesian grid based on a one-continuum formulation where the two phases are distinguished using an indicator function. 

The time integration method used to solve equations (\ref{eq:NS}), (\ref{eq:B-adv}), and  (\ref{eq:PHI-adv}) is based on a explicit fractional-step method \citep{kim_moin_1985a}, where all the terms are advanced with the third order Runge-Kutta scheme, except the solid hyper-elastic contribution which is advanced with Crank-Nicolson \citep{min_yoo_choi_2001a}. At each time step the volume fraction $\phi^s$ and left Cauchy-Green deformation tensor $B_{ij}$ are updated first
\begin{subequations}
\label{eq:adv-num}
\begin{align}
\label{eq:PHI-adv-num}
\frac{\phi^{k}-\phi^{k-1}}{\Delta t} &+ \beta^{k} \left(  \frac{\partial u_k \phi}{\partial x_k} \right)^{k-1} + \gamma^{k} \left(  \frac{\partial u_k \phi}{\partial x_k} \right)^{k-2} = 0, \\
\label{eq:B-adv-num}
\begin{split}
\frac{B_{ij}^{k}-B_{ij}^{k-1}}{\Delta t} &+ \beta^{k} \left(  \frac{\partial u_k B_{ij}}{\partial x_k} - B_{kj}\frac{\partial u_i}{\partial x_k} - B_{ik}\frac{\partial u_j}{\partial x_k}\right)^{k-1} + \\
&+\gamma^{k} \left(  \frac{\partial u_k B_{ij}}{\partial x_k} - B_{kj}\frac{\partial u_i}{\partial x_k} - B_{ik}\frac{\partial u_j}{\partial x_k} \right)^{k-2} = 0,
\end{split}
\end{align}
\end{subequations}
followed by the prediction step of the momentum conservation equations
\begin{equation}
\label{eq:NS-num1}
\begin{split}
\frac{u_i^{*}-u_i^{k-1}}{\Delta t} &+ 2 \alpha^{k} \frac{1}{\rho} \frac{\partial p^{(k-1)}}{\partial x_i} + \alpha^{k} \left( - \frac{1}{\rho} \frac{\partial \sigma^{sh}_{ij}}{\partial x_j} \right)^{k} + \alpha^{k} \left( - \frac{1}{\rho} \frac{\partial \sigma^{sh}_{ij}}{\partial x_j} \right)^{k-1} + \\
&+ \beta^{k} \left(  \frac{\partial u_i u_j}{\partial x_j} - \frac{1}{\rho} \frac{\partial \sigma'_{ij}}{\partial x_j} \right)^{k-1} + \gamma^{k} \left(  \frac{\partial u_i u_j}{\partial x_j} - \frac{1}{\rho} \frac{\partial \sigma'_{ij}}{\partial x_j}  \right)^{k-2} = 0.
\end{split}
\end{equation}
In the previous equations, $\Delta t$ is the overall time step from $t^n$ to $t^{n+1}$, the superscript $^*$ is used for the predicted velocity, while the superscript $^k$ denotes the Runge-Kutta substep, with $k=0$ and $k=3$ corresponding to time $n$ and $n+1$. 

The pressure equation that enforces the solenoidal condition on the velocity field is solved via a Fast Poisson Solver
\begin{equation}
\label{eq:NS-num2}
\frac{\partial \psi ^{k}}{\partial x_i \partial x_i} = \frac{\rho}{2 \alpha^{k} \Delta t} \frac{\partial u_i^{*}}{\partial x_i},
\end{equation}
and, finally, the pressure and velocity are corrected according to
\begin{subequations}
\label{eq:NS-num3}
\begin{align}
p^{k} &= p^{k-1} + \psi ^{k}, \\
u_i^{k} &= u_i^{*} - 2 \alpha^{k} \frac{\Delta t}{\rho} \frac{\partial \psi ^{k} }{\partial x_i},
\end{align}
\end{subequations}
where $\psi$ is the projection variable, and $\sigma_{ij}'$ is the deviatoric stress tensor. $\alpha$, $\beta$, and $\gamma$ are the integration constants, whose values are
\begin{equation}
\label{eq:RK}
\begin{array}{ccc}
\alpha^1 = \frac{4}{15} & \alpha^2 = \frac{1}{15} & \alpha^3 = \frac{1}{6}\\ \\
\beta^1 = \frac{8}{15} & \beta^2 = \frac{5}{12} & \beta^3 = \frac{3}{4}\\ \\
\gamma^1 = 0 & \gamma^2 = -\frac{17}{60} & \gamma^3 = -\frac{5}{12}
\end{array}
\end{equation}
The governing differential equations are solved on a staggered grid using a second order central finite-difference scheme, except for the advection terms in \equref{eq:adv-num} where the fifth-order WENO scheme is applied, as it was proved to work properly by \citet{sugiyama_ii_takeuchi_takagi_matsumoto_2011a}. A comprehensive review on the properties of different numerical schemes for the advection terms is reported by \citet{min_yoo_choi_2001a}.

The Zalesak's disk \citep{zalesak_1979a}, \ie a slotted disc undergoing solid body rotation, is a standard benchmark to validate the numerical schemes for advection problems, since the initial shape should not deform under solid body rotation. The set-up is the same described by \citet{zalesak_1979a} and the comparison of the initial shape (black line) and that after one full rotation (red and blue lines) is shown in \figrefA{fig:zalesak}. Two grid resolutions are considered here, $64$ and $128$ grid points per diameter, shown in blue and red, respectively. The final shape of the disk shows an overall good agreement with the initial one, with  better agreement on the finer grid. As expected, the major differences are located on the edges of the geometry.

\begin{figure}
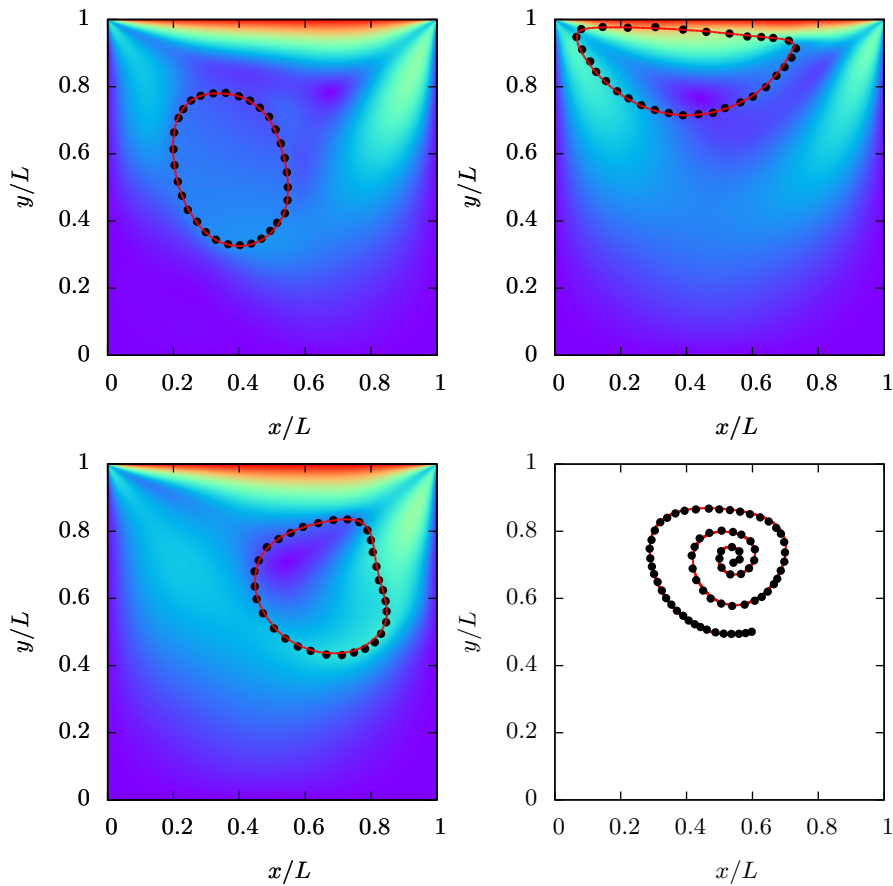

  \centering
  \input{fig3a}
  \input{fig3b}
  \vspace{0.8cm} \\
  \input{fig3c}
  \input{fig3d}
  \vspace{0.5cm}
  \caption{(a)-(c) Comparison of the hyper-elastic solid deformation in the lid-driven cavity with published results. The solid line represents our results, while the dots are taken from \citet{sugiyama_ii_takeuchi_takagi_matsumoto_2011a}. The different snapshots correspond to the time $t=2.34L/V_w$, $4.69L/V_w$, and $7.03L/V_w$. The color background contour is the magnitude of the velocity, and the color scale goes from $0$ (violet) to $1V_w$ (red). (d) Trajectories of the solid centroid in the lid-driven cavity in a time range $t\in\left[0,20L/V_w\right]$.}
  \label{fig:val}
\end{figure}

\subsection{Code validation} \label{sec:validation}
The code has been validated by performing a full $2D$ Eulerian simulation of a deformable solid motion in a lid-driven square cavity of side $L$, at a Reynolds number of $Re=\rho V_w L/\mu=100$, where $V_w$ is the velocity of the moving top wall, and $\rho$ and $\mu$ the density and dynamic viscosity of both phases (\ie $\rho=\rho^s=\rho^f$ and $\mu=\mu^s=\mu^f$). The system is initially at rest, and the initial solid shape is circular with a radius of $0.2L$, centered at $\left( 0.6L, 0.5L \right)$. The solid phase is neo-Hookean material with modulus of transverse elasticity $G=0.1\rho V_w^2$. The numerical domain is discretised with $512\times512$ grid points.

\figrefC[a-c]{fig:val} visualizes the solid deformation at three different equispaced time instants, starting from $t=1.17L/V_w$ to $t=7.03L/V_w$. The solid lines represents the instantaneous particle shapes, while the dots are the results by \citet{sugiyama_ii_takeuchi_takagi_matsumoto_2011a}. Further comparison is provided in \figref[d]{fig:val}, where we display the trajectory of the solid centroid for $t\in\left[0,20L/V_w\right]$ from our simulation (solid line) and that provided by \citet{sugiyama_ii_takeuchi_takagi_matsumoto_2011a}. The particle shape and trajectory is in very good agreement with the published results. Note that, a comparison between this fully Eulerian procedure and a Lagrangian one is provided in \citet{sugiyama_ii_takeuchi_takagi_matsumoto_2011a}, together with other validations and test cases.

\subsection{Numerical details} 
For all the turbulent flows considered hereafter, the equations of motion are discretised by using $1296 \times 540 \times 648$ grid points on a computational domain of size $6h \times 2.5h \times 3h$ in the streamwise, wall-normal and spanwise directions. The spatial resolution has been chosen in order to properly resolve the wall deformation, satisfying the constraint $\Delta x^{w+} = \Delta y^{w+} = \Delta z^{w+} < 0.8$. For the case at highest Reynolds number, a grid refinement study was performed using $1944 \times 810 \times 972$ grid points in the streamwise, wall-normal and spanwise directions ($50\%$ more in each direction). The difference in the resulting friction coefficient $C_f$ results to be less than $2\%$.

All the simulations are started from a fully developed turbulent channel flow with a perfectly flat interface between the fluid and solid phases. After the flow has reached statistical steady state, the calculations are continued for an interval of $600 h/U_b$ time units, during which $120$ full flow fields are stored for further statistical analysis. To verify the convergence of the statistics, we have computed them using different number of samples and verified that the differences are negligible.

\section{Results} \label{sec:result}
\begin{figure}
  \centering
  \input{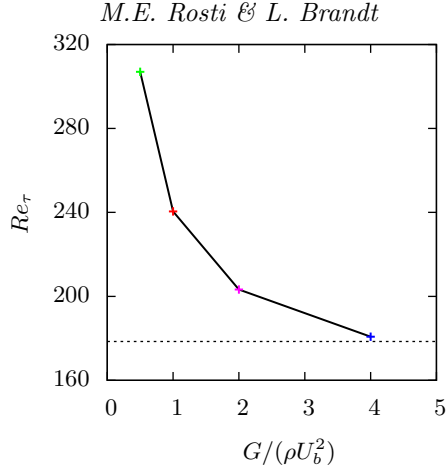}
  \vspace{0.5cm}
  \caption{Friction Reynolds number of the top deformable wall $Re_\tau$, as a function of the elastic modulus $G$. The coloured symbols blue, magenta, red, and green are used for the cases of decreasing elastic modulus $G$. The dashed horizontal line in the figure represents the value for the rigid wall case \citep{kim_moin_moser_1987a}.}
  \label{fig:reynolds}
\end{figure}
We study turbulent channel flows over viscous hyper-elastic walls, together with a baseline case over stationary impermeable walls taken by the seminal work of \citet{kim_moin_moser_1987a}. All the simulations are performed at constant flow rate, so that the flow Reynolds number based on the bulk velocity is fixed at $2800$, \ie $Re=\rho U_b h/\mu^f=2800$, where the bulk velocity is the average value of the mean velocity computed across the whole domain occupied by the fluid phase. This choice facilitates the comparison between the flow in a channel with elastic walls and the flow in a channel bounded by rigid impermeable walls. In general, if the pressure gradient is kept constant in time, the flow rate oscillates in time around a constant value. On the other hand, if the flow rate is kept constant in time the pressure gradient oscillates around a constant value. In the present work, consistently with choosing $U_b$ as the characteristic velocity, we opt for enforcing the constant flow rate condition; hence the appropriate value for the instantaneous value of the streamwise pressure gradient is determined at every time step.

The elastic wall in the unstressed condition is flat and parallel to the bottom rigid wall, and its reference height is fixed to $h_e=0.5h$. The modulus of transverse elasticity $G$ is varied and four values are considered here, ranging from an almost rigid case, $G=2.0 \rho U_b^2$, to a highly elastic one, $G=0.25 \rho U_b^2$. The viscosity of the solid is set equal to the fluid one, \ie $\mu^s=\mu^f=\mu$, except for the results in the last section, section \ref{sec:result-solidvisc}, where we report data for two solid materials with larger and lower viscosities. In particular, the viscosity ratio $\mu^s/\mu^f$ is set equal to $0.1$ and $10$. The full set of simulations is reported in \tabref{tab:cases}, together with  some mean quantities, such as the resulting friction Reynolds number based on the bottom rigid wall $Re_\tau^w$ and based on the top elastic wall $Re_\tau$, the maximum velocity $\overline{u}_M$, and its $y$-coordinate location $y_M$. The friction Reynolds number $Re_\tau$ is also shown in \figrefA{fig:reynolds} as a function of the elastic modulus $G$.

\begin{table}
\centering
\setlength{\tabcolsep}{5pt}
\begin{tabular}{lrrrrrrr}
Case													&	$h_e/h$	&	$\mu^s/\mu^f	$	&	$G/\left( \rho U_b^2 \right)$	&	$Re_\tau^w$	&	$Re_\tau$	&	$\overline{u}_M/U_b$		&	$y_M/h$		\\
\hline
Reference											&	$-$		&	$-$				&	$-$							&	$178.5$		&	$178.5$		&	$1.16$					&	$0.000$		\\
$G\downarrow$										&	$0.5$	&	$1$				&	$4.0$						&	$179.0$		&	$180.8$		&	$1.17$					&	$-0.030$		\\
$G\downarrow\downarrow$								&	$0.5$	&	$1$				&	$2.0$						&	$182.1$		&	$203.3$		&	$1.19$					&	$-0.076$		\\
$G\downarrow\downarrow\downarrow$					&	$0.5$	&	$1$				&	$1.0$						&	$193.0$		&	$240.5$		&	$1.23$					&	$-0.206$		\\
$G\downarrow\downarrow\downarrow\downarrow$			&	$0.5$	&	$1$				&	$0.5$						&	$202.3$		&	$307.0$		&	$1.32$					&	$-0.386$		\\
$\mu^s\downarrow$									&	$0.5$	&	$0.1$			&	$1.0$						&	$192.9$		&	$240.4$		&	$1.22$					&	$-0.197$		\\
$\mu^s\uparrow$										&	$0.5$	&	$10.$			&	$1.0$						&	$187.0$		&	$226.0$		&	$1.19$					&	$-0.127$		\\
\end{tabular}
\caption{Summary of the DNSs performed at different viscous hyper-elastic parameters, all at a fixed bulk Reynolds number equal to $Re=2800$.}
\label{tab:cases}
\end{table}

Viscous units, used above to express spatial resolution, will be often employed in the following; they are indicated by the superscript $^+$, and are built using the friction velocity $u_\tau$ as velocity scale and the viscous length $\delta_\nu = \nu / u_\tau$ as length scale. For a turbulent channel flow with solid walls, the dimensionless friction velocity is defined as
\begin{equation} 
\label{eq:friction_velocity}
u_\tau = \sqrt{\dfrac{1}{Re} \left. \dfrac{d \overline{u}}{dy} \right\vert_{y=0}},
\end{equation}
where $\overline{u}$ is the mean velocity, and the subscript indicates that the derivative is taken at $y=0$, the location of the solid wall. When the channel has moving walls, the definition \eqref{eq:friction_velocity} must be modified to account for the turbulent shear stresses that are in general non-zero at the solid-fluid interface. Similarly to what usually done for porous walls \citep{breugem_boersma_uittenbogaard_2006a}, we define
\begin{equation} \label{eq:friction_velocity_total}
u_\tau = \sqrt{\dfrac{1}{Re} \left. \dfrac{d \overline{u}}{d y} \right\vert_{y=2} - \left. \overline{u'v'} \right\vert_{y=2}},
\end{equation}
where $\overline{u'v'}$ is the off-diagonal component of the Reynolds stress tensor and the quantities are evaluated at the mean interface location, $y=2$. In our simulations two viscous units can be defined, one based on the bottom rigid wall, and one on the top elastic wall, by using  \equref{eq:friction_velocity} and \equref{eq:friction_velocity_total}. In the following, we will distinguish the two units adding a superscript $^w$ when referring to the bottom rigid wall. The actual value of the friction velocity of the elastic wall is computed from its friction coefficient, found by combining the information of the total $C_f$, obtained from the driving streamwise pressure gradient, and the one of the lower rigid wall found by \equref{eq:friction_velocity}.

\subsection{Mean flow}
\figrefC[a]{fig:mean-vel} shows the mean velocity profile $\overline{u}$ as a function of the wall-normal coordinate. The black $+$ symbol  is used for the reference case with rigid walls taken from \citet{kim_moin_moser_1987a}, and reproduced with the same code in \cite{picano_breugem_brandt_2015a};  the solid lines indicate the results from the present simulations with blue, magenta, red, and green used for cases with increasing elasticity (lower transverse elastic modulus $G$). This color scheme will be used through the whole paper. The presence of one rigid and one elastic wall makes the mean velocity profile skewed, with its maximum $\overline{u}_M$ increasing and located closer to the bottom wall as the elasticity increases. Note that the mean velocity of the elastic wall is equal to zero.
\begin{figure}
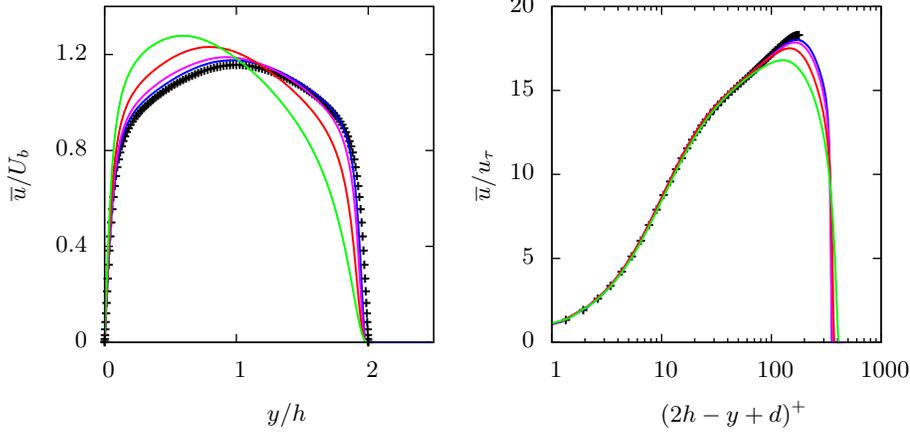

  \centering
  \input{fig5a}
  \input{fig5b}
  \vspace{0.5cm}
  \caption{(a) Comparison of the streamwise mean velocity profile $U$ of a turbulent channel flow at $Re=2800$ over rigid walls ($+$ symbols) and over different hyper-elastic walls (solid lines). The blue, magenta, red, and green lines are used for the cases $G\downarrow$, $G\downarrow\downarrow$, $G\downarrow\downarrow\downarrow$, and $G\downarrow\downarrow\downarrow\downarrow$, respectively, \ie for increasing elasticity (lower transverse elastic modulus $G$). The results for the rigid case are taken from \citet{kim_moin_moser_1987a}. (b) Mean velocity profile $U$ versus the distance from the bottom rigid wall in wall units.}
  \label{fig:mean-vel}
\end{figure}

\figrefC[b]{fig:mean-vel} shows the mean velocity profiles versus the logarithm of the distance from the bottom rigid wall in wall-units. For the turbulent channel flows over rigid impermeable walls (shown with the symbol $+$), we can usually identify three regions: firstly, the  viscous sublayer for $y^{w+}<5$ where the variation of $\overline{u}^{w+}$ with $y^{w+}$ is approximately linear, \ie
\begin{equation} \label{eq:viscous-sublayer}
\overline{u}^{w+}=y^{w+}.
\end{equation}
In the so-called log-law region, $y^{w+}>30$ the variation of $\overline{u}^{w+}$ versus $y^{w+}$ is logarithmic, \ie
\begin{equation} \label{eq:log-law}
\overline{u}^{w+}=\dfrac{1}{k} \log y^{w+} + B,
\end{equation}
defined by the coefficients $k$ (the von Karman constant) and $B$. The values of $k$ and $B$ for smooth rigid walls are usually assumed to be $k=0.41$ and $B=5$. However, for the Reynolds number that we are considering, a better fit with experimental and numerical simulations is obtained with $k=0.40$ and $B=5.5$. Finally, the region between $5$ and $30$ wall units is called buffer layer and neither laws hold.

As shown in \figref[b]{fig:mean-vel}, the velocity profiles from the cases with an elastic wall still coincide, which indicates that the scaling of the mean flow near the bottom wall is not altered by the presence of the deformable top wall, the only differences being a reduced inertial range and the location of the maximum velocity at lower wall-distances due to the skewness of the mean velocity profile.
\begin{table}
\centering
\setlength{\tabcolsep}{8pt}
\begin{tabular}{lrrrrr}
Case													&	$d/h$	&	$k+\Delta k$		&	$\Delta U^+$	&	$\delta/h$	&	$\delta'/h$	\\
\hline
Reference											&	$-$		&	$0.41$	&	$-$				&	$-$			&	$-$			\\
$G\downarrow$										&	$0.011$	&	$0.36$	&	$1.5$		&	$0.011$		&	$0.003$		\\
$G\downarrow\downarrow$								&	$0.069$	&	$0.29$	&	$7.5$		&	$0.061$		&	$0.007$		\\
$G\downarrow\downarrow\downarrow$					&	$0.120$	&	$0.25$	&	$14.5$		&	$0.095$		&	$0.025$		\\
$G\downarrow\downarrow\downarrow\downarrow$			&	$0.193$	&	$0.20$	&	$24.4$		&	$0.126$		&	$0.042$		\\
$\mu^s\downarrow$									&	$0.120$	&	$0.25$	&	$14.5$		&	$0.090$		&	$0.015$		\\
$\mu^s\uparrow$										&	$0.055$	&	$0.31$	&	$7.9$		&	$0.054$		&	$0.008$		\\
\end{tabular}
\caption{Summary of the log-law fitting parameters and wall deformations for the cases shown in \tabref{tab:cases}. The table provides the origin shift $d$, the modified von Karman constant $k+\Delta k$, the shift of the velocity profile $\Delta U^+$, the maximum and root mean square values of the wall deformation $\delta$ and $\delta'$.}
\label{tab:cases-loglaw}
\end{table}

\begin{figure}
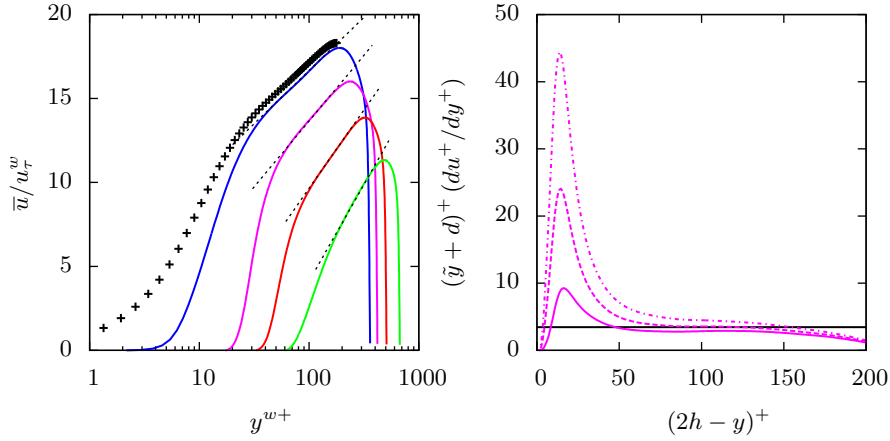

  \centering
  \input{fig6a}
  \input{fig6b}
  \vspace{0.5cm}
  \caption{(a) Mean velocity profile $U$ versus the distance from the top rigid wall $\tilde{y}+d$ in wall units, and (b) $(\tilde{y} + d)^+ d\overline{u}^+/d\tilde{y}^+$ as a function of $\tilde{y}^+$ for the case $G\downarrow\downarrow$. The three lines in (b) correspond to three values of $d$, \ie $0.00$ (solid line), $0.07$ (dashed line), and $0.14$ (dash-dotted line). The colour scheme is the same as in \figrefA{fig:mean-vel}.}
  \label{fig:mean-vel-log}
\end{figure}

The mean velocity profiles in inner units based on the friction at the top moving walls are shown in \figref[a]{fig:mean-vel-log}. Differently from \figref[b]{fig:mean-vel}, the profiles are plotted versus $\tilde{y}+d=(2h-y)+d$, where $d$ is a shift of the origin. This shift was introduced by \citet{jackson_1981a} for rough walls and then used by \cite{breugem_boersma_uittenbogaard_2006a} for permeable walls.  Introducing $d$, the log-law is modified as
\begin{equation} \label{eq:log-law-jackson1}
\overline{u}^{+}=\dfrac{1}{\lambda} \log \frac{\tilde{y}+d}{y_0},
\end{equation}
where $\lambda$ is the slope of the inertial range, which may differ from $k$, \ie $\lambda=k+\Delta k$, and $y_0$ is the so-called equivalent roughness height. Through simple manipulation, the previous relation can be rewritten as
\begin{equation} \label{eq:log-law-jackson2}
\overline{u}^{+}=\dfrac{1}{\lambda} \log \left( \tilde{y} + d \right)^+ - \frac{1}{\lambda} \log y_0^+,
\end{equation}
and for a better comparison with \equref{eq:log-law} as
\begin{equation} \label{eq:log-law-mod}
\overline{u}^{+}=\dfrac{1}{k+\Delta k} \log \left( \tilde{y} + d \right)^+ + B - \Delta U^+.
\end{equation}

In order to identify the value of $d$, we have first plot $(\tilde{y} + d)^+ d\overline{u}^+/d\tilde{y}^+$ as a function of $\tilde{y}^+$ for several values of $d$. Inside the logarithmic layer this quantity must be constant and equal to the inverse of the slope in the inertial range, \ie $1/\left( k+\Delta k \right)$. In this way, the correct value of the shift can be identified as the one providing a region with constant $(\tilde{y} + d)^+ d\overline{u}^+/d\tilde{y}^+$ (an equilibrium range). The procedure is exemplified in \figref[b]{fig:mean-vel-log} where we report $(\tilde{y} + d)^+ d\overline{u}^+/d\tilde{y}^+$  for the case $G\downarrow\downarrow$ and three values of $d$, \ie $0.00$ (solid line), $0.07$ (dashed line), and $0.14$ (dash-dotted line). Note that the slope in the inertial range is slightly positive  for $d=0.0$, while  it becomes negative for $d=0.14$; only for the correct value $d=0.07$ the slope is null, providing $k+\Delta k=0.294$. Once $d$ and $\Delta k$ are known, $\Delta U^+$ can be found by a simple fitting.

This procedure has been repeated for all the cases, obtaining the values of the fitting parameters reported in \tabref{tab:cases-loglaw} and used for the scaling in \figref[a]{fig:mean-vel-log}; note that the dashed lines reported in the figure are computed using these values. All the cases with elastic walls under consideration show an increase in the slope of the logarithmic region, $\Delta k<0$, when compared with the flow over rigid walls, as well as a downward shift $\Delta U^+$. The change of slope in the logarithmic region, corresponding to an increase of drag, was also found in turbulent channel flow over porous walls by \citet{breugem_boersma_uittenbogaard_2006a} for high values of permeabilities, while no change was found for low values of permeability in \cite{rosti_cortelezzi_quadrio_2015a} and for the turbulent channel flow over rough walls by \citet{orlandi_leonardi_2008a}; indeed the only effect of a rough wall was a downward shift of the mean profile, also associated to a drag increase.

As discussed by \citet{orlandi_leonardi_2008a}, previous studies employing Direct Numerical Simulations with different boundary conditions for the three velocity components at the wall have shown that the turbulent flow is strongly linked to the fluctuations of the wall-normal velocity component $v'$, with rms values denoted by a $'$ in the following. This has been shown by \cite {orlandi_leonardi_tuzi_antonia_2003a} for the case of a rough surface made of two-dimensional square bars that can be mimicked by distributions of $v'$, by \citet{flores_jimenez_2006a} who used synthetic velocity distributions, and by \citet{jimenez_uhlmann_pinelli_kawahara_2001a} who modelled a permeable wall adding a transpiration boundary condition proportional to the local pressure fluctuations $p'$, \ie $v=-\beta p'$, $\beta$ being a constant. Also, \citet{cheng_castro_2002a} experimentally showed that the roughness function correlates well with $v'$. Based on these previous data, \citet{orlandi_leonardi_2006a, orlandi_leonardi_2008a} proposed first a correlation between $\Delta U^+$ and $v'$, then improved by using $v'^+$, finally showing that
\begin{equation} \label{eq:deltau-roughness}
\Delta U^+ \approx \frac{B}{k} v'^+.
\end{equation}
\begin{figure}
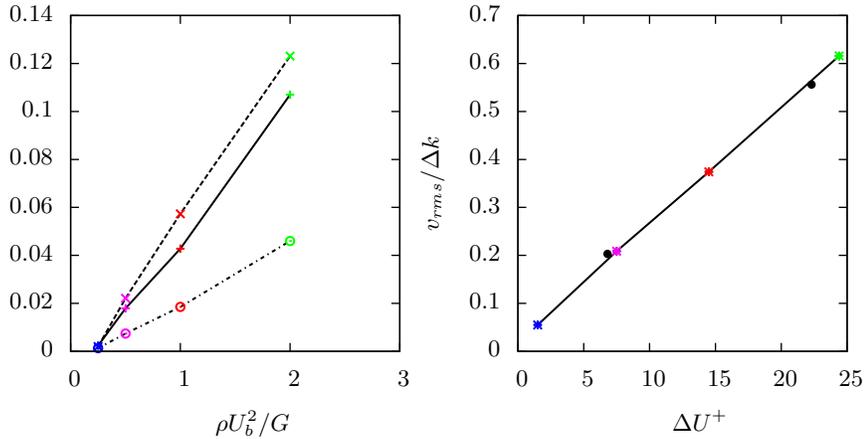

  \centering
  \input{fig7a}
  \input{fig7b}
  \vspace{0.5cm}
  \caption{(a) Rms values of velocity fluctuations at the interface, as function of the inverse of the elastic modulus ($\rho U_b^2/G$). The solid line is used for the streamwise component, the dashed line for the wall-normal component, and the dash-dotted one for the spanwise component. (b) Rms of the wall-normal velocity component $v'$ versus the velocity shift $\Delta U^+$, see \equref{eq:log-law-v}. The black circles are the data for the turbulent channel flow over porous wall investigated by \citet{breugem_boersma_uittenbogaard_2006a}. The colour scheme is the same as in \figrefA{fig:mean-vel}.}
  \label{fig:rms-vel-log}
\end{figure}
These correlation was validated against the results by \cite {orlandi_leonardi_tuzi_antonia_2003a, orlandi_leonardi_2006a, orlandi_leonardi_2008a, flores_jimenez_2006a}, and \citet{cheng_castro_2002a}. Both numerical and experimental data fit well with relationship \eqref{eq:deltau-roughness} for values of $v'^+$ up to $0.8$; above this values the linear correlation departs from the data, showing hence discrepancies at high $v'^+$. 

In order to test a similar correlation also for the present configuration, we display the rms of the velocity fluctuation components at the mean interface location, $y=2h$, versus the elasticity $\rho U_b^2/G$ in \figref[a]{fig:rms-vel-log}. In the picture, the solid, dashed and dash-dotted lines are used for the streamwise, wall-normal and spanwise velocity components. All the three components of the fluctuating velocity appear to vary almost linearly with the inverse of the elasticity, the wall-normal component of the velocity being the fastest and the spanwise the slowest. \figrefC[b]{fig:rms-vel-log} shows the log-law shift $\Delta U^+$ as a function of the ratio between the wall-normal velocity fluctuation $v'$ and the change of the inertial range slope $\Delta k$. For all the cases considered, we find that a linear relation well approximates the two quantities, \ie
\begin{equation} \label{eq:log-law-v}
\Delta U^+ \approx \frac{v'}{\Delta k}.
\end{equation}
The proposed fit, given in \equref{eq:log-law-v}, is here tested also for the case of a turbulent channel flow over porous wall by using the data of the simulations in \citet{breugem_boersma_uittenbogaard_2006a}, shown in \figref[b]{fig:rms-vel-log} as black circles. Interestingly, the agreement is quite good also in this case  (please note, that the values of $\Delta U^+$ and $v'$ are not reported but extracted from the figures).

\subsection{Velocity fluctuations}
\begin{figure}
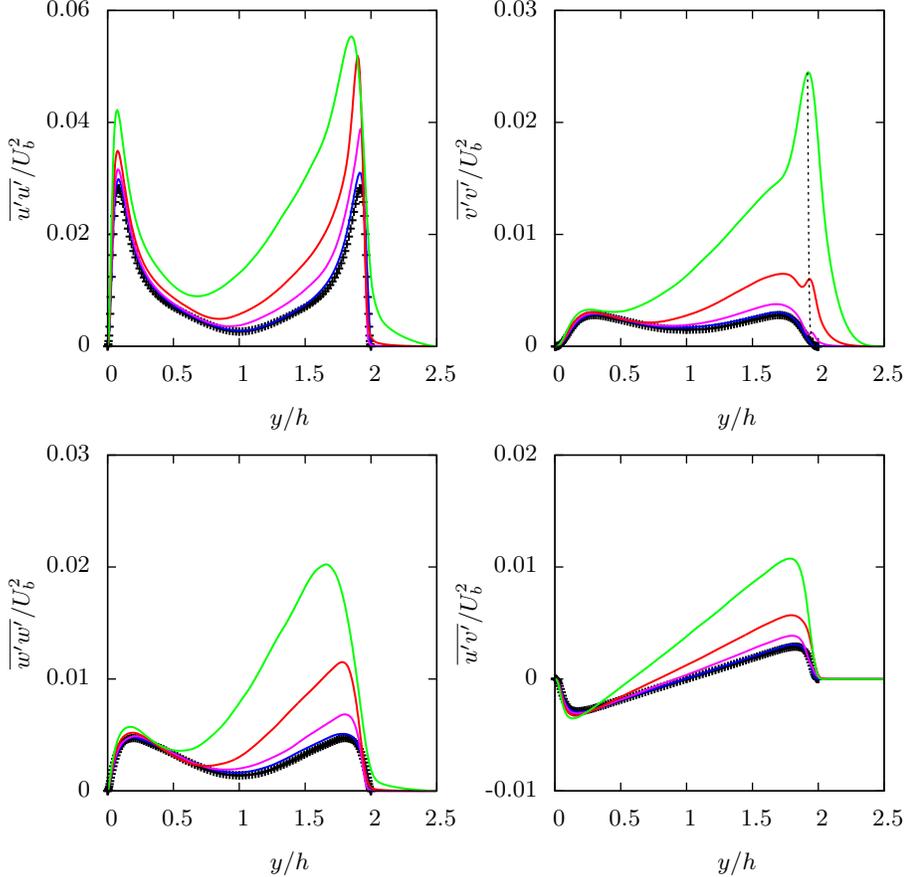

  \centering
  \input{fig8a}
  \input{fig8b}
  \vspace{0.8cm} \\
  \input{fig8c}
  \input{fig8d}
  \vspace{0.5cm}
  \caption{Wall-normal profiles of the different components of the Reynolds stress tensor, normalised with $U_b^2$ for a turbulent channel flow over hyper-elastic walls. The symbols indicate the flow between two rigid walls  from \citet{kim_moin_moser_1987a}. Panels (a), (b), and (c) show the diagonal components $u'u'$, $v'v'$, and $w'w'$, while (d) the cross term $u'v'$. The colour scheme is the same as in \figrefA{fig:mean-vel}.}
  \label{fig:rms}
\end{figure}
We continue our comparison between the turbulent channel flows with rigid and elastic walls by analyzing the wall-normal distribution of the diagonal component of the Reynolds stress tensor; these are shown in \ref{fig:rms}(a-c) together with the data from \citet{kim_moin_moser_1987a} for the rigid case. All the components are strongly affected by the moving wall, and the effect is not limited to the region close to the interface, but it extends also beyond the centreline. The peaks of the Reynolds stresses move farther from the walls as the elasticity is increased. As already showed in \figref[a]{fig:rms-vel-log}, all the components have non-zero values at $y=2h$ for the elastic cases, since the no-slip condition is now enforced on a wall which is moving, \ie $u_i^f=u_i^s$. The Reynolds stress components decrease inside the solid layer, eventually becoming zero at the rigid top wall, located at $y=2.5h$. Note that for the most elastic case considered ($G\downarrow\downarrow\downarrow\downarrow$) all the stress components, especially the wall-normal one, do not clearly vanish until reaching the rigid top wall, thus indicating that the fluctuations propagate deeply inside the solid layer. In relative terms, the spanwise and wall-normal components are the most affected by the wall elasticity, which can be attributed to a weakening of the wall-blocking and wall-induced viscous effects \citep{perot_moin_1995a, perot_moin_1995b}, and observed also in the experiments of \citet{krogstad_antonia_browne_1992a} and \cite{krogstadt_antonia_1999a} on boundary-layer flows over rough walls. The wall-normal component displays a secondary peak very close to the interface, which grows with the elasticity and eventually becomes its maximum value. This secondary peak is associated with the oscillatory movement of the wall. Indeed, the dashed line in  \figref[b]{fig:rms} connects the  maximum of the wall-normal oscillation amplitude $\delta/h$ reported in \tabref{tab:cases-loglaw} with the local value of the Reynolds stress, thus, clearly showing that the location of the peak coincides with the maximum oscillation of the wall. We also note that the peak of the streamwise component of the fluctuating velocity, $u'^2$, grows much less then that pertaining the other two components. This increase is even lower than the growth of the friction Reynolds number, corresponding to a reduction in wall units. The reduced amplitude of the streamwise fluctuations is usually associated with the absence or reduction of streaky structures above a wall \citep{rosti_cortelezzi_quadrio_2015a, breugem_boersma_uittenbogaard_2006a}, which is true also in the present case, as shown in the next section.

Finally, \figref[d]{fig:rms} depicts the wall-normal profile of the off-diagonal component of the Reynolds stress tensor. Also this cross component is strongly affected by the presence of the moving wall. In particular, the maximum value increases and moves away from the wall as the elasticity increases. Consequently, also the location where the stress is zero $y_{uv}$, \ie zero turbulent production, moves closer to the bottom solid walls, from $y=1h$ for the rigid case to $y\approx 0.58h$ for the most elastic upper wall. Nevertheless the stress profiles vary linearly between the two peaks, of opposite sign, close to each wall, although with different slopes. At the interface the value of the stress is not null as in the rigid case, however, inside the elastic layer the stress vanishes quickly. 

\begin{figure}
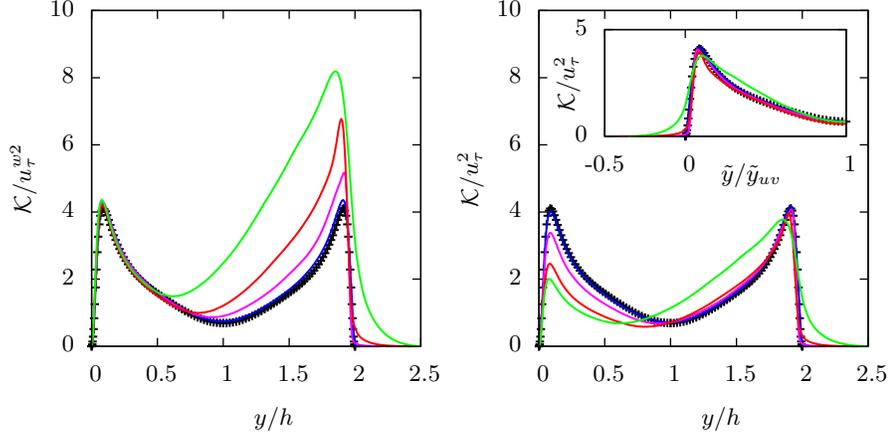

  \centering
  \input{fig9a}
  \input{fig9b}
  \vspace{0.5cm}
  \caption{Wall-normal profiles of the turbulent kinetic energy $\mathcal{K}=\left( u'^2 + v'^2 + w'^2 \right)/2$, normalised with the friction velocities of (a) the bottom rigid wall $u_\tau^w$ and (b) of the top elastic wall $u_\tau$. Symbols indicate the flow between two rigid walls from the data in  \citet{kim_moin_moser_1987a}. The colour scheme is the same as in \figrefA{fig:mean-vel}.}
  \label{fig:kin}
\end{figure}

\begin{figure}
  \centering
  \includegraphics[width=0.49\textwidth]{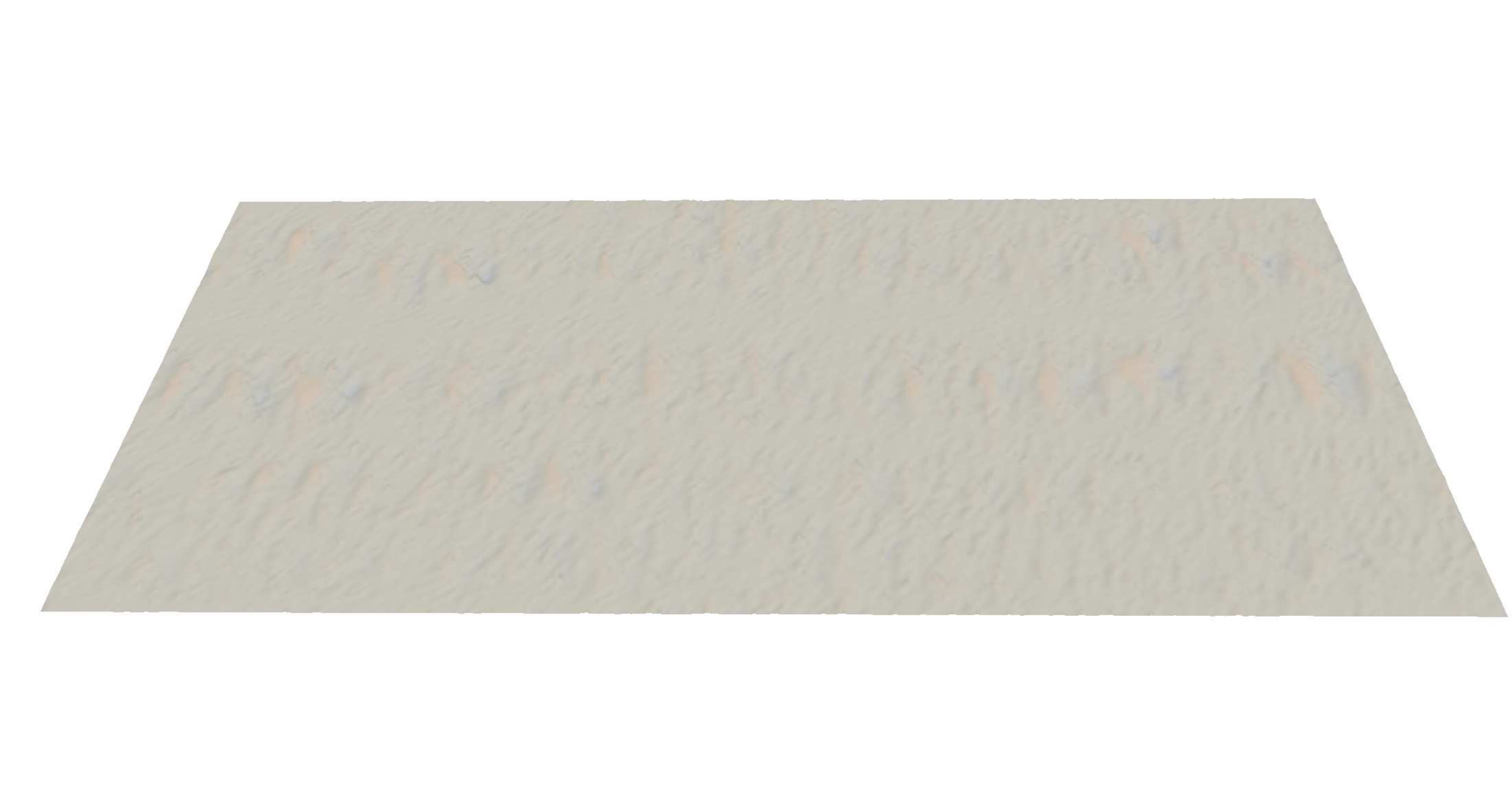}
  \includegraphics[width=0.49\textwidth]{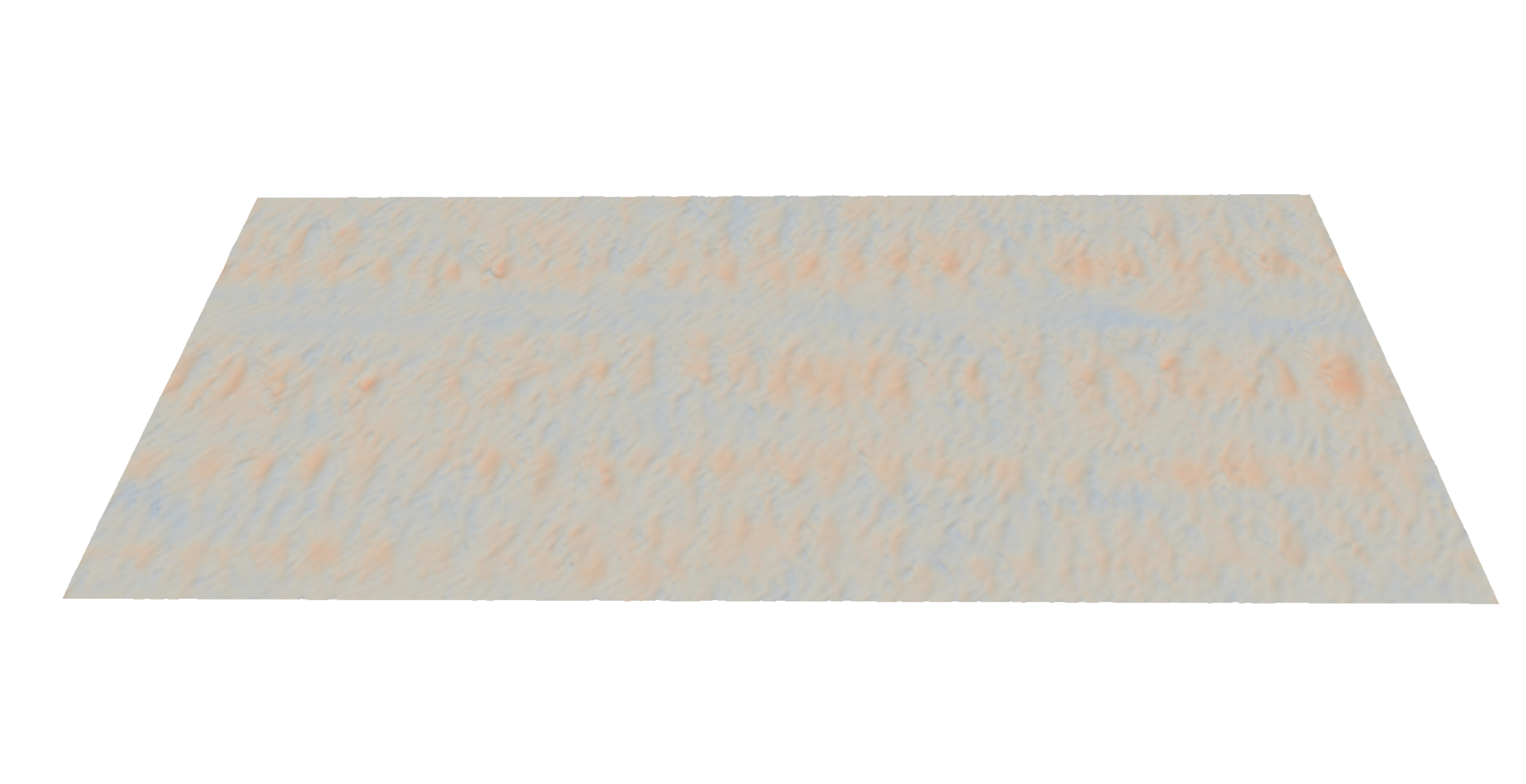}
  \includegraphics[width=0.49\textwidth]{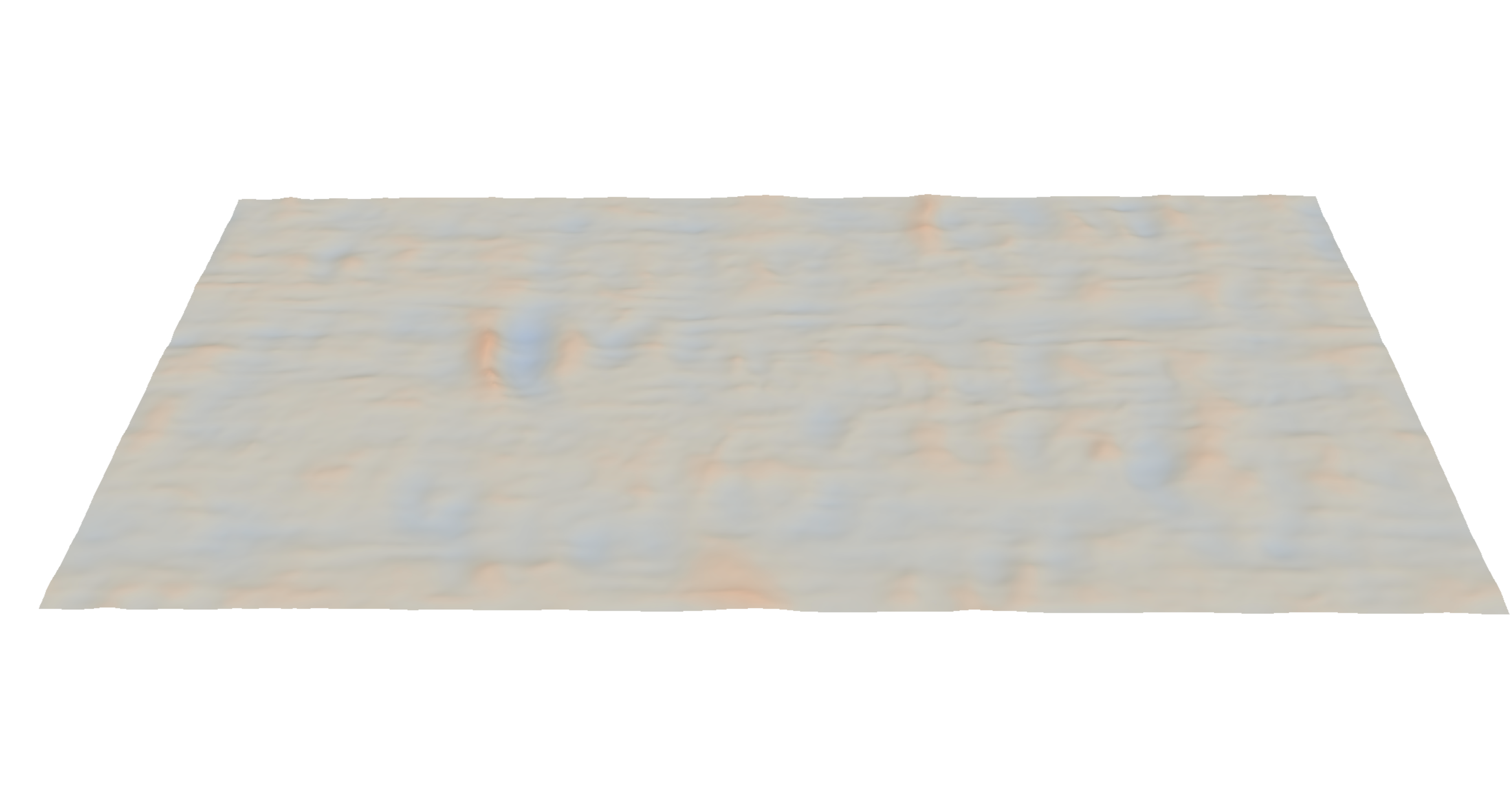}
  \includegraphics[width=0.49\textwidth]{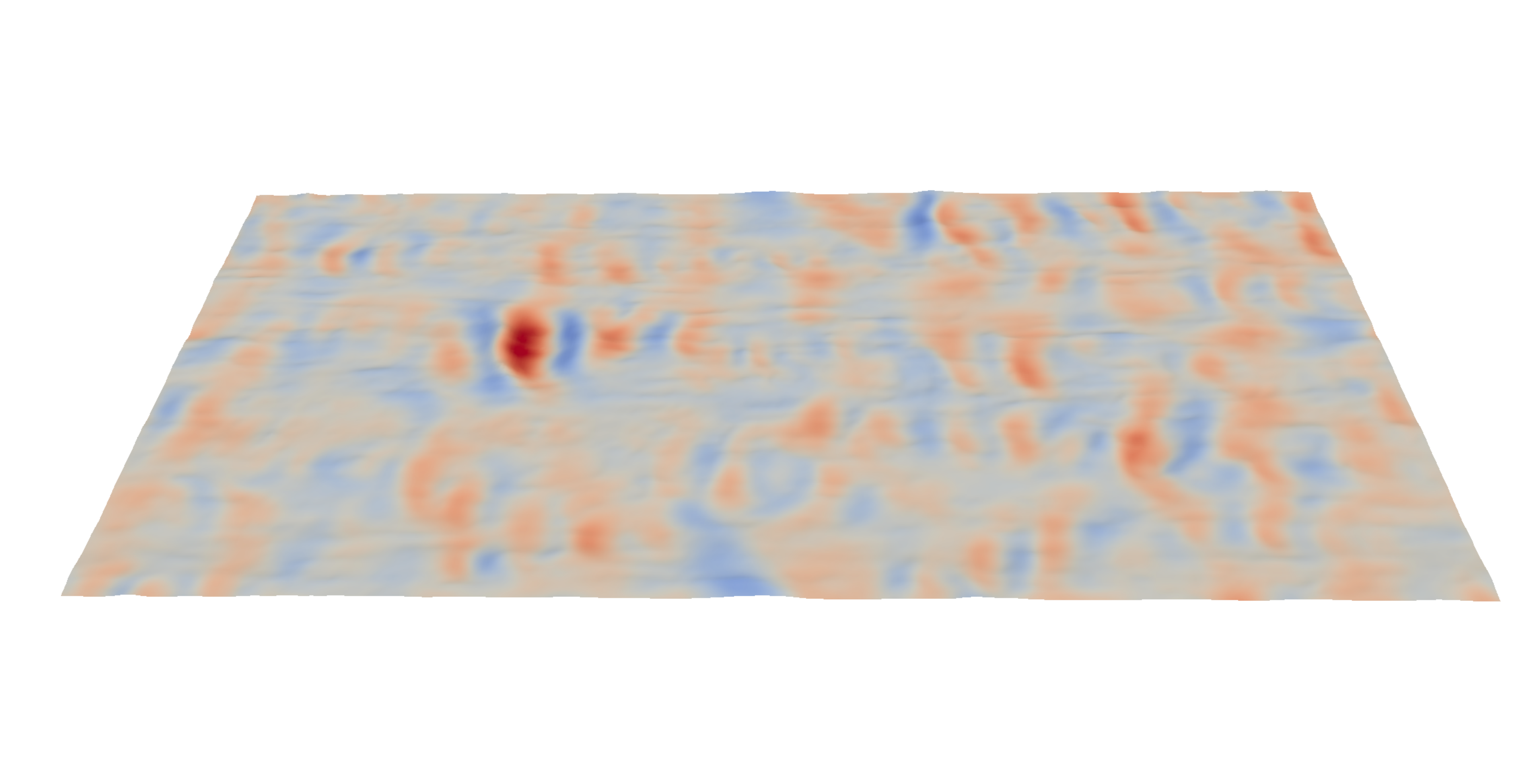}
  \includegraphics[width=0.49\textwidth]{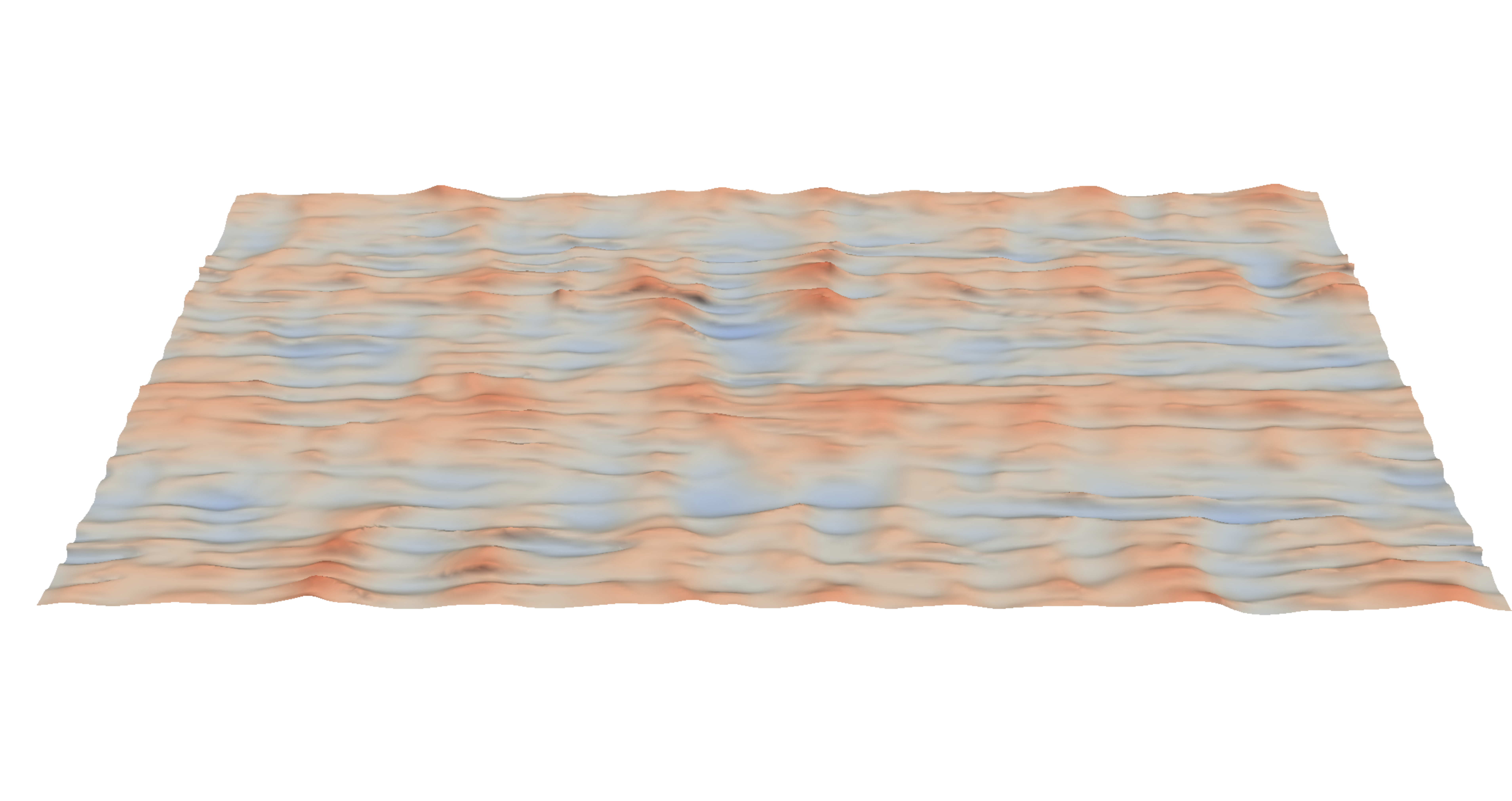}
  \includegraphics[width=0.49\textwidth]{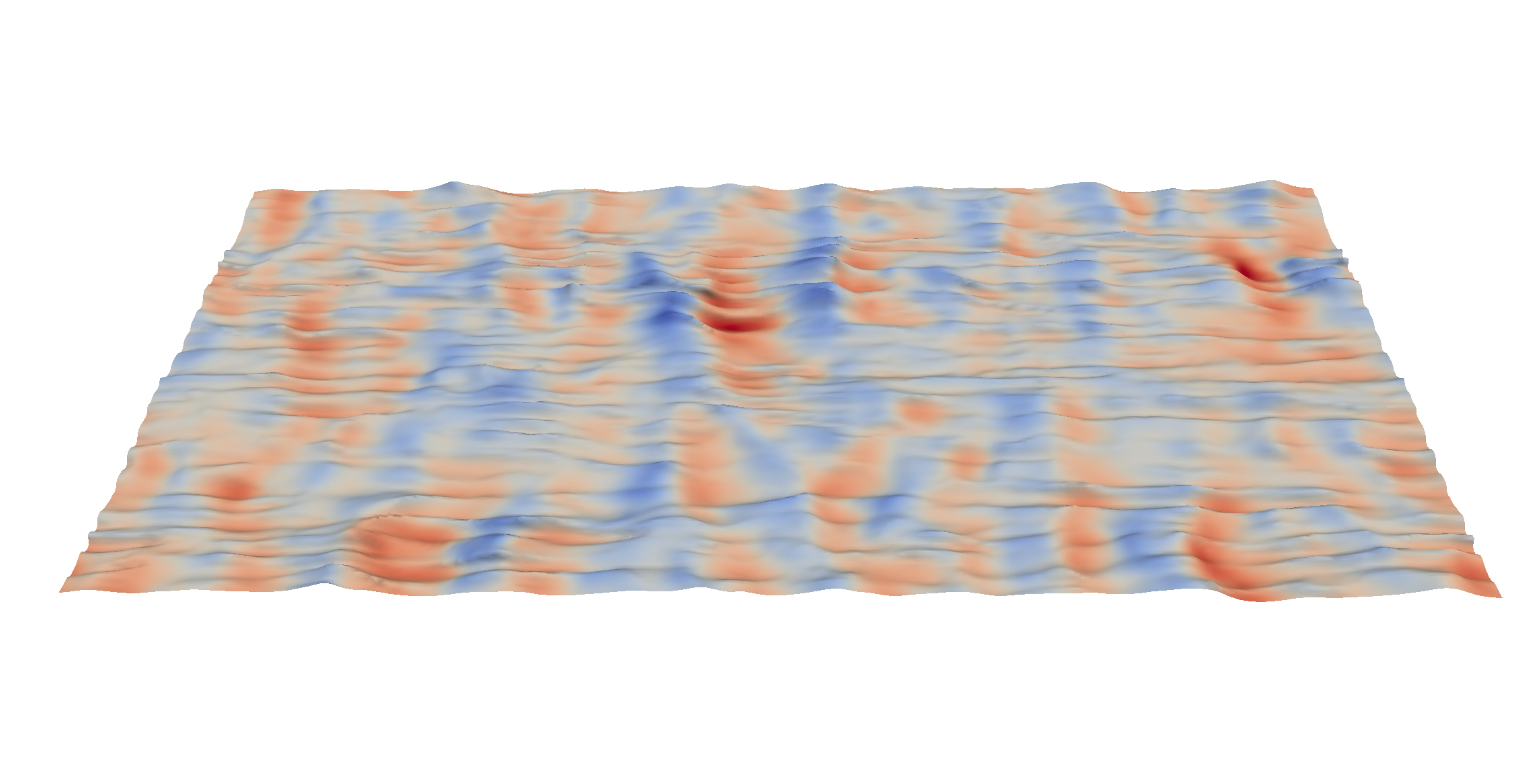}
  \includegraphics[width=0.49\textwidth]{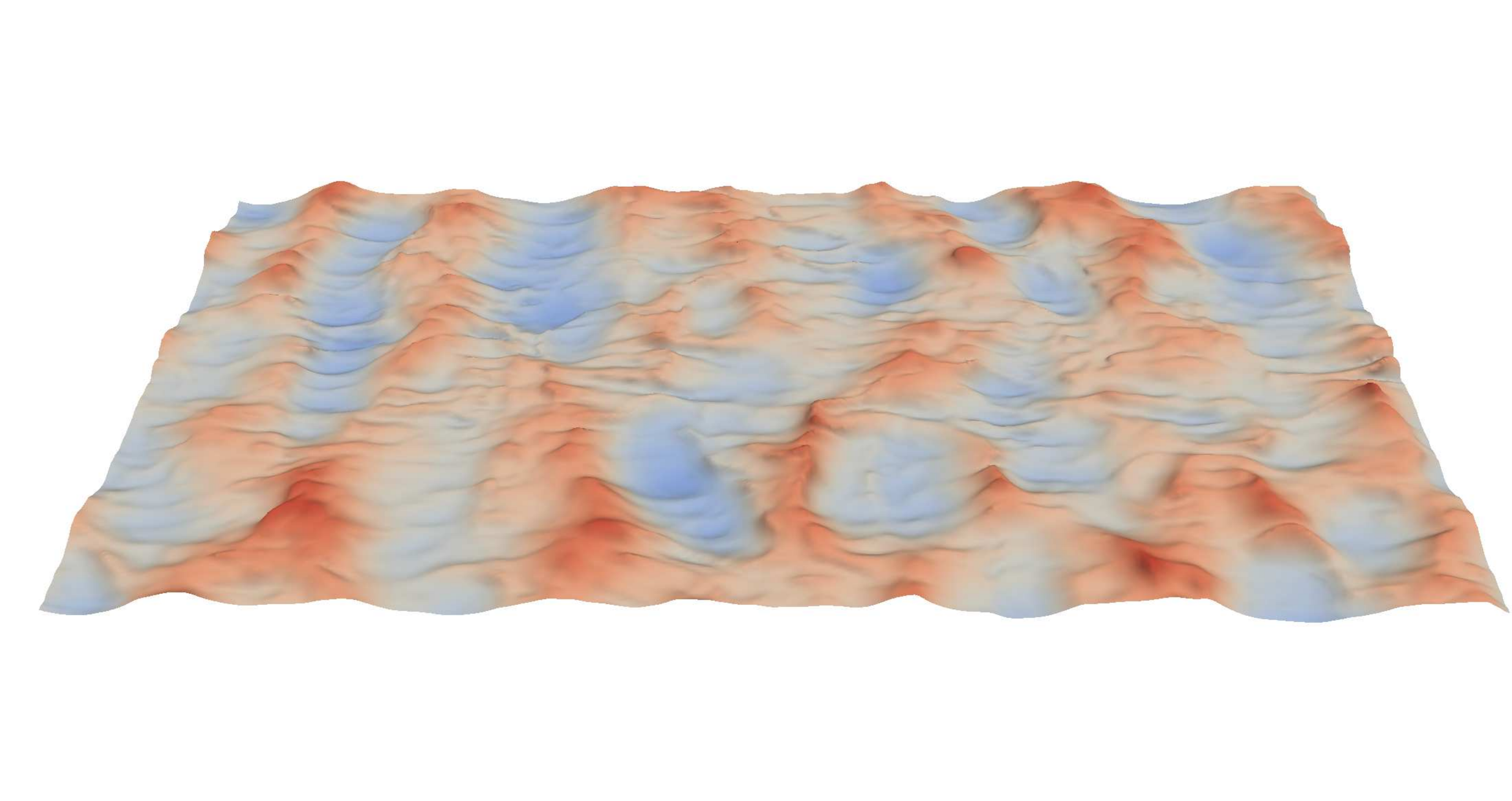}
  \includegraphics[width=0.49\textwidth]{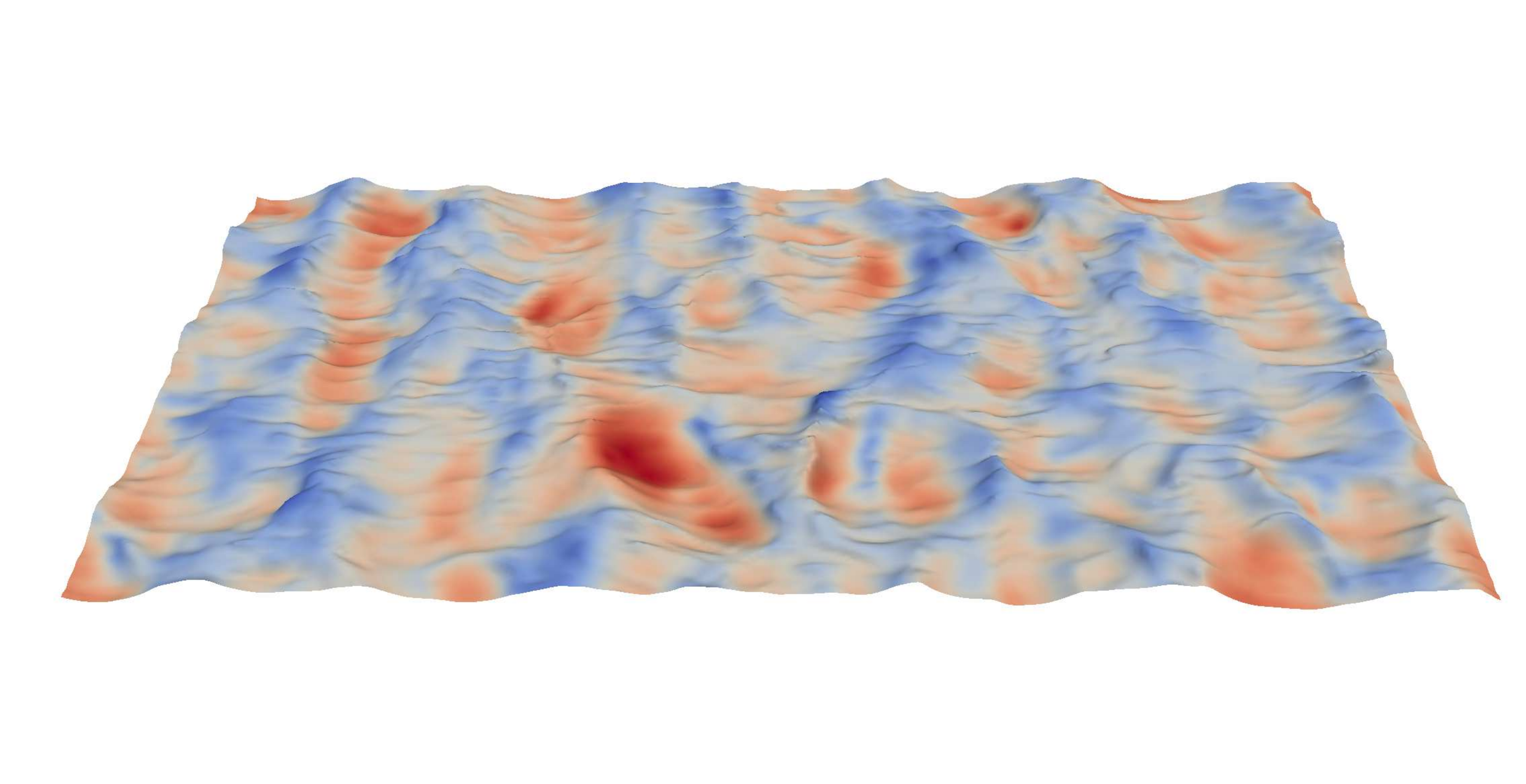}
  \includegraphics[width=0.24\textwidth]{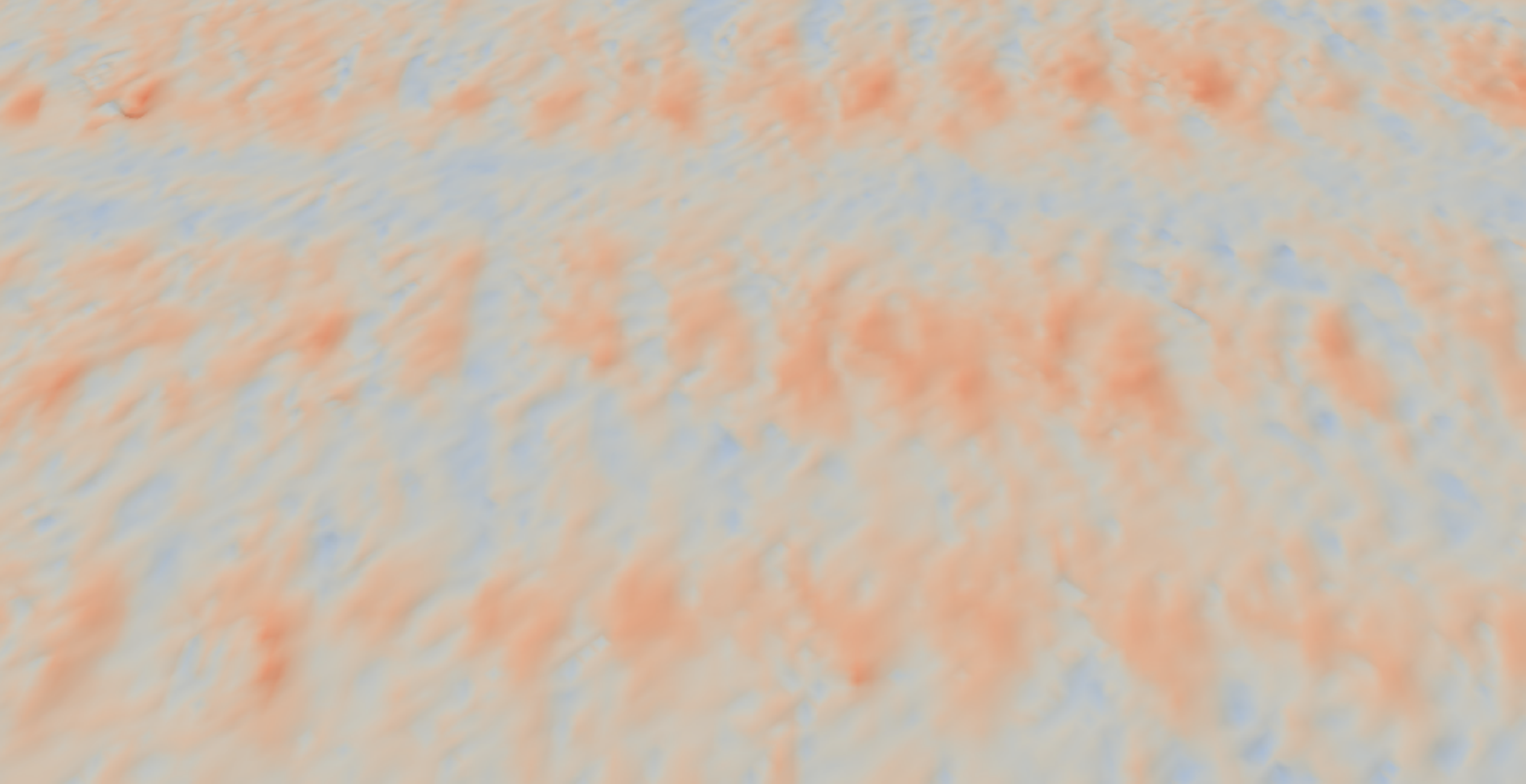}
  \includegraphics[width=0.24\textwidth]{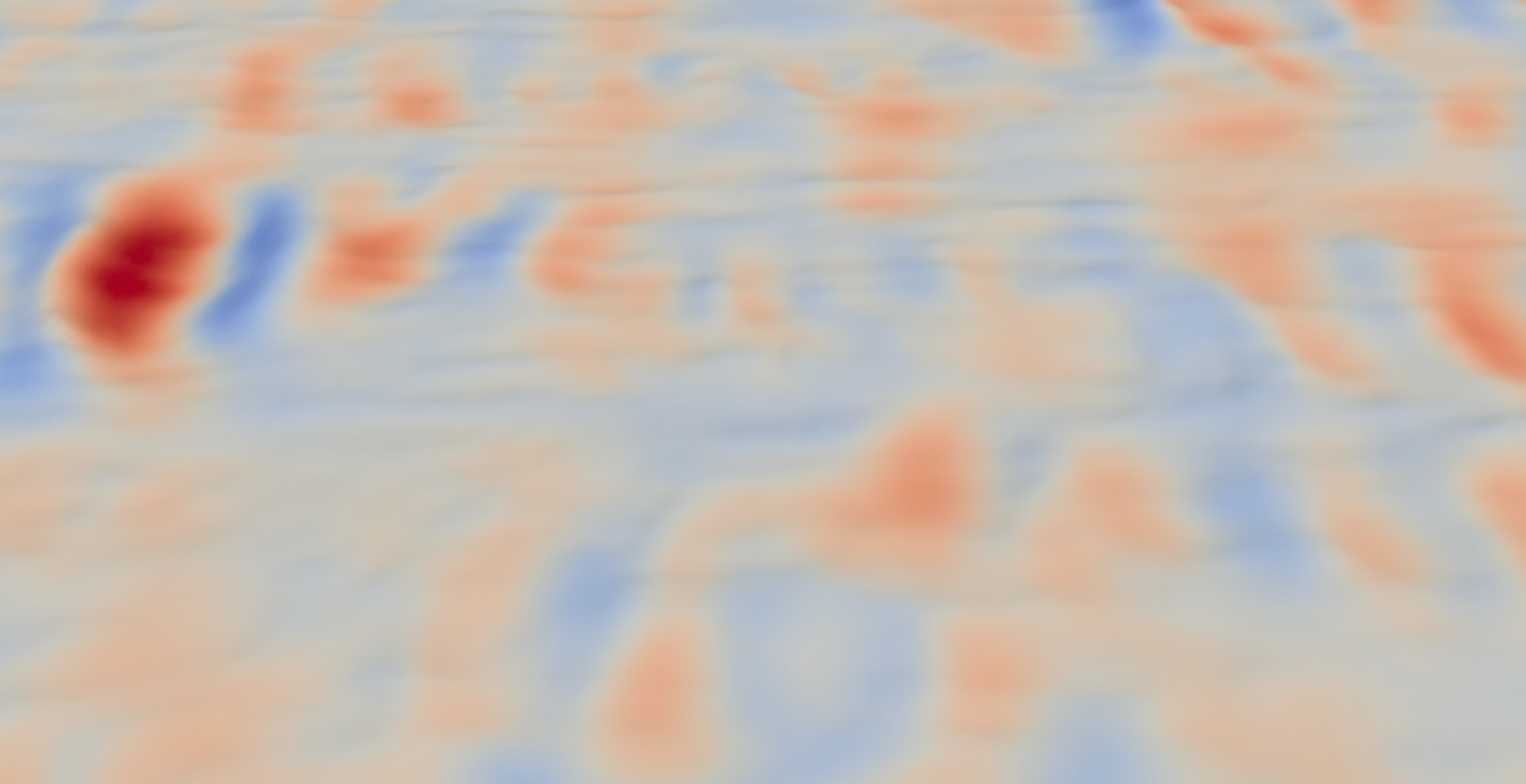}
  \includegraphics[width=0.24\textwidth]{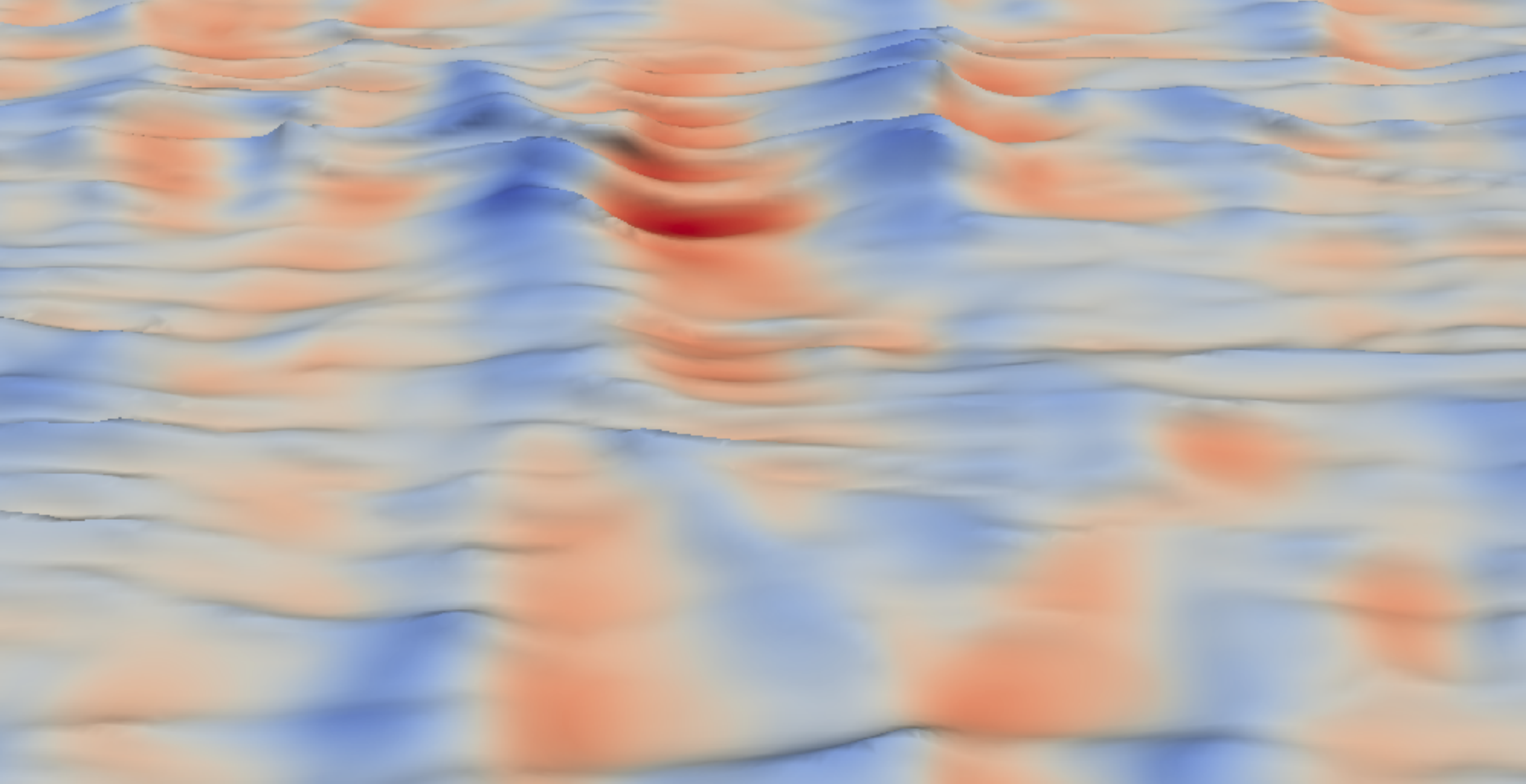}
  \includegraphics[width=0.24\textwidth]{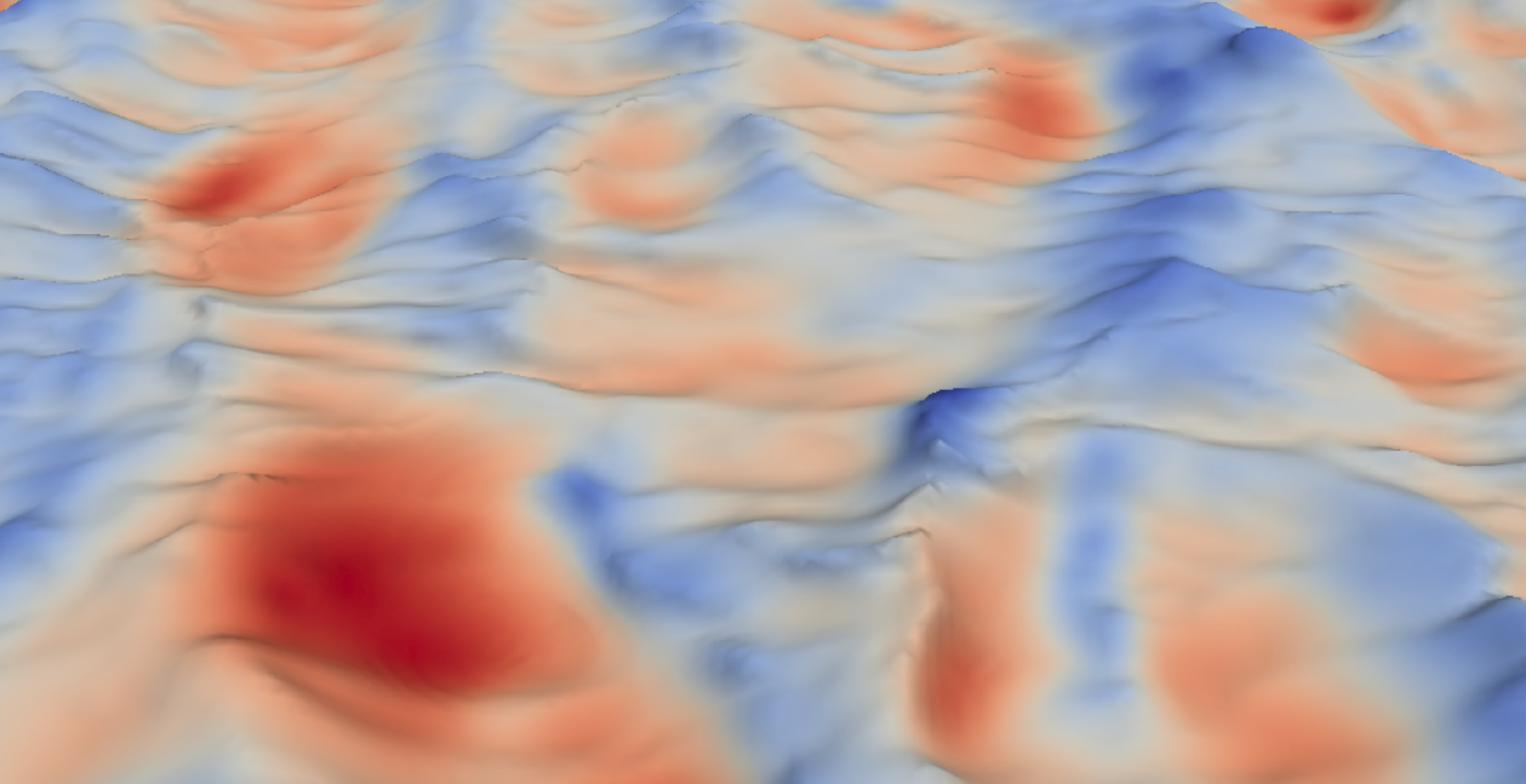}
  \caption{Instantaneous configuration of the top elastic wall, with transverse elastic modulus $G$ decreasing from top to bottom. The flow goes from left to right. The surfaces are coloured by the wall-normal distance on the left column, with the color scale ranging from $-0.13h$ (blue) to $0.13h$ (red), and by the pressure on the right column, with the color scale ranging from $-1 \rho U_b^2$ (blue) to $1 \rho U_b^2$ (red). The bottom row is a zoomed view of pictures on the left column.}
  \label{fig:moving-wall}
\end{figure}

An overall view of the velocity fluctuations can be inferred considering the turbulent kinetic energy $\mathcal{K}=\left( u'^2 + v'^2 + w'^2 \right)/2$, shown in \figrefA{fig:kin}. We display in \figref[a]{fig:kin}  the kinetic energy normalised by the friction velocity $u_\tau^w$ at the bottom rigid wall as a function of the wall-normal distance $y/h$. As usual, the symbols represent the profiles from the DNS of \citet{kim_moin_moser_1987a} of turbulent flow between two solid rigid walls. Close to the bottom wall the profiles coincide, as expected, which indicates that the influence of the moving wall on the velocity fluctuations near the other wall is negligible. In the boundary layer close to the moving wall, instead, we observe a strong increase of the kinetic energy, with the maximum value becoming the double of the peak close to the bottom wall. \figrefC[b]{fig:kin} reports the same turbulent kinetic energy profiles, but now normalized by the friction velocity $u_\tau$ at the moving top wall. Except for the case $G\downarrow\downarrow\downarrow\downarrow$, the profiles tend to coincide away from the wall (for $y$ in the range between $1h$ and $1.8h$), whereas the peak close to the moving wall is slightly lower for higher wall elasticities. The reduced peak is mainly due to the weak increase (decrease in wall units) of the streamwise component of the velocity fluctuations discussed above. The inset in \figref[b]{fig:kin} shows again the turbulent kinetic energy profile close to the moving wall versus the wall normal distance $\tilde{y}$ divided by the coordinate $\tilde{y}_{uv}$ where the shear component of the Reynolds stress tensor cross zero. This scaling compensates for the asymmetry in the geometry, similarly to what done by \citet{leonardi_orlandi_antonia_2007a}, and provides a better collapse of the profiles.

\subsection{Solid deformation and flow structures}
We now discuss in more details the deformation of the wall and its effect on the emerging flow coherent structures. \figrefAC{fig:moving-wall} shows one instantaneous configuration of the top elastic wall for the four cases considered here with growing elasticity from top to bottom, and with the flow from left to right. The wall surface is coloured by the wall-normal distance to the reference straight unstressed condition on the left column, with the color scale ranging from $-0.13h$ (blue) to $0.13h$ (red), and by the pressure on the right column, with the color scale ranging from $-1 \rho U_b^2$ (blue) to $1 \rho U_b^2$ (red). The wall surface deforms with a maximum amplitude of the wall deformation $\delta$ ranging approximately from $1\%$ of the channel half height $h$ for the most rigid case ($G\downarrow$) to $13\%$ in the most deformable one ($G\downarrow\downarrow\downarrow\downarrow$). We recall that the values of $\delta$ are reported in \tabref{tab:cases-loglaw}, together with the root mean square value of the wall deformation $\delta'$. It is interesting to note that the spanwise coherency of the wall deformation tends to increase as $G$ is decreased, \ie the deformation increases.

\begin{figure}
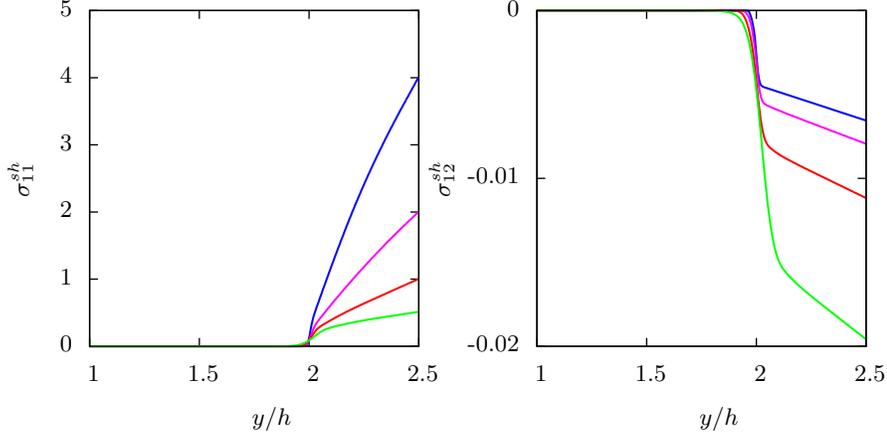

  \centering
  \input{fig11a}
  \input{fig11b}
  \vspace{0.5cm}
  \caption{Profiles of the (a) streamwise $\sigma_{11}^{sh}$ and (b) cross $\sigma_{12}^{sh}$ hyper-elastic stress tensor components as a function of the wall-normal coordinate $y/h$ for the different values of the transverse elastic modulus $G$ under investigation. The colour scheme is the same as in \figrefA{fig:mean-vel}.}
  \label{fig:stress}
\end{figure}

\figrefAC{fig:stress} shows the mean profile of the hyper-elastic stress tensor streamwise $\sigma_{11}^{sh}$ and shear $\sigma_{12}^{sh}$ components (see \equref{eq:stress-s-he}). We note that the mean streamwise component is zero in the fully fluid region ($y>2h-\delta$) and rapidly grows for $2h-\delta<y<2h+\delta$, before reaching its positive maximum value at the solid wall, $y=2.5h$. The shear component follows a similar trend, but has a maximum negative value at the wall, with the shear component being one or two order of magnitude smaller then the streamwise one. While $\sigma_{11}^{sh}$ decreases as $G$ is reduced, \ie for higher wall elasticity, on the contrary, $\sigma_{12}^{sh}$ increases (in absolute value). Note also that the values of the stress components approximately scale with $G$.

\begin{figure}
  \centering
  \includegraphics[width=0.45\textwidth]{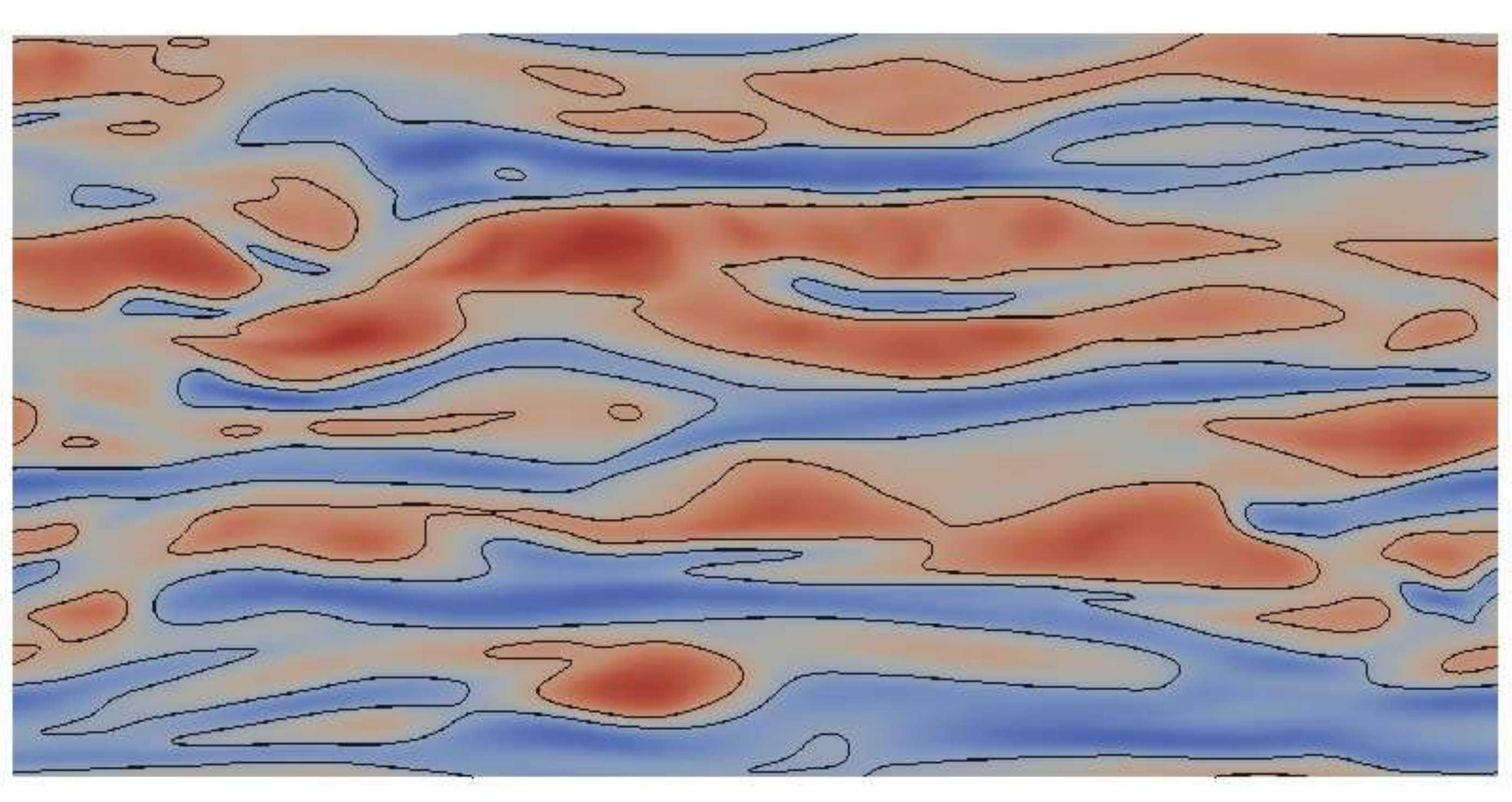}
  \includegraphics[width=0.45\textwidth]{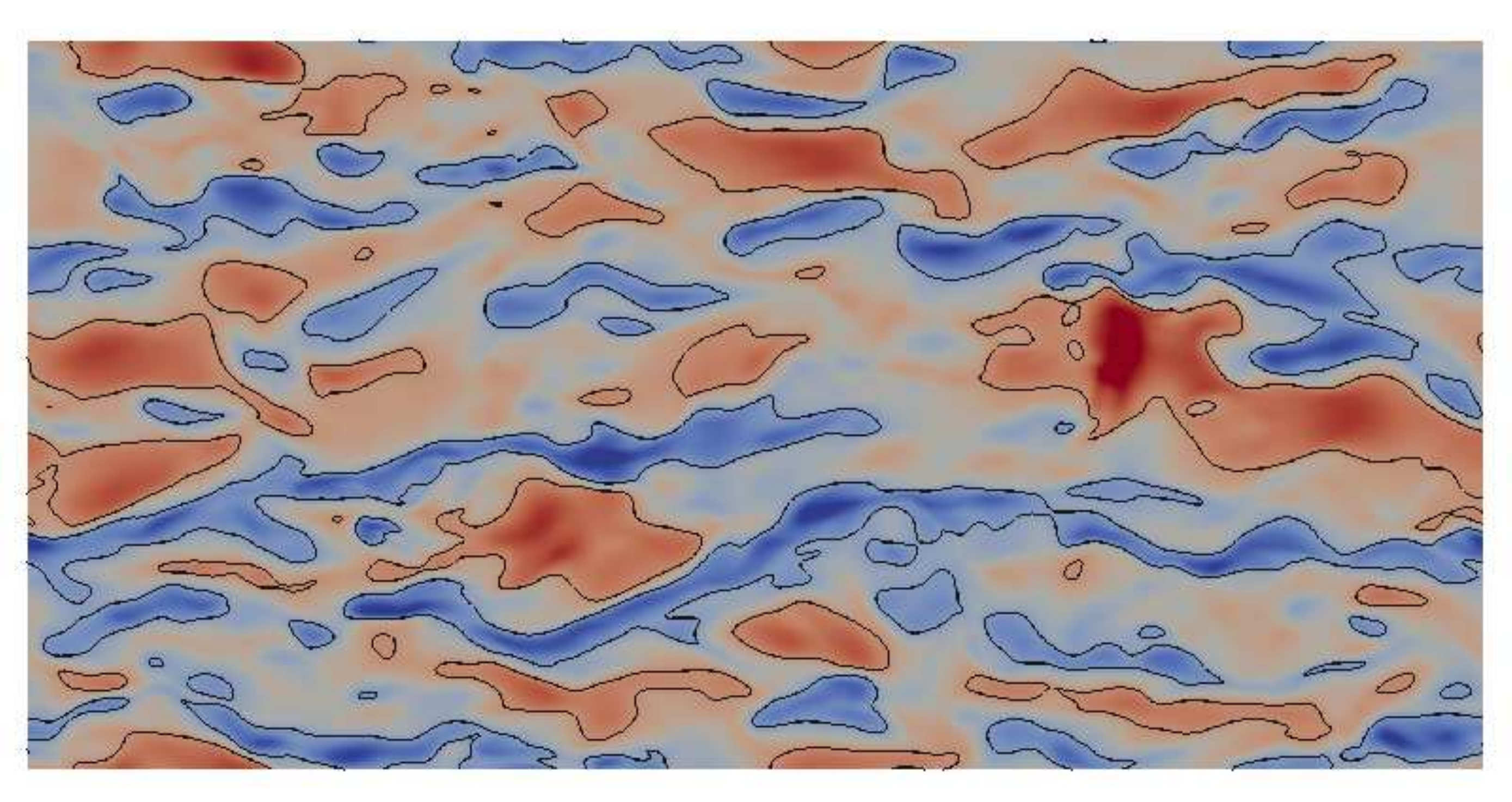} \\ \vspace{0.8mm}
  \includegraphics[width=0.45\textwidth]{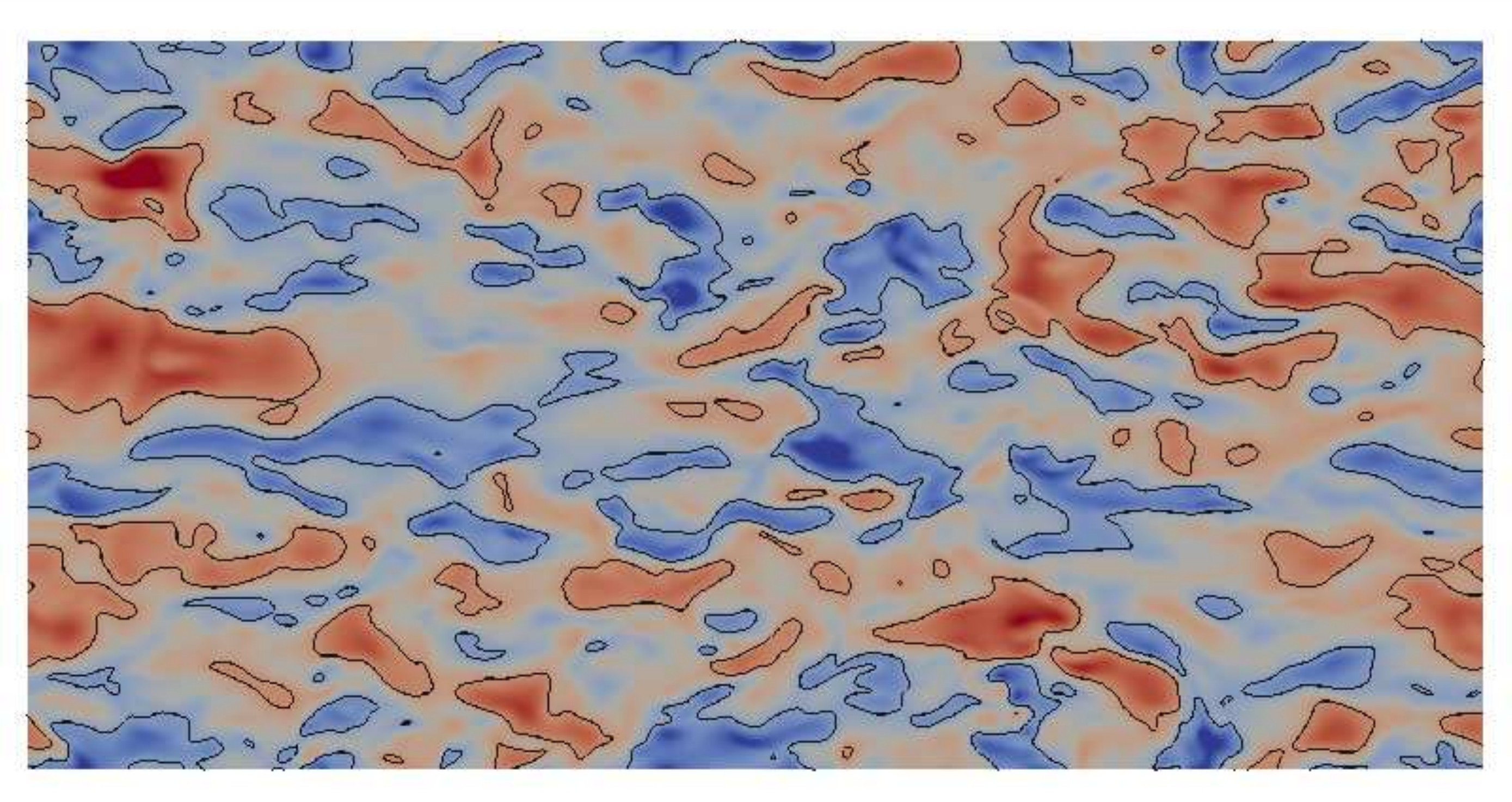}
  \includegraphics[width=0.45\textwidth]{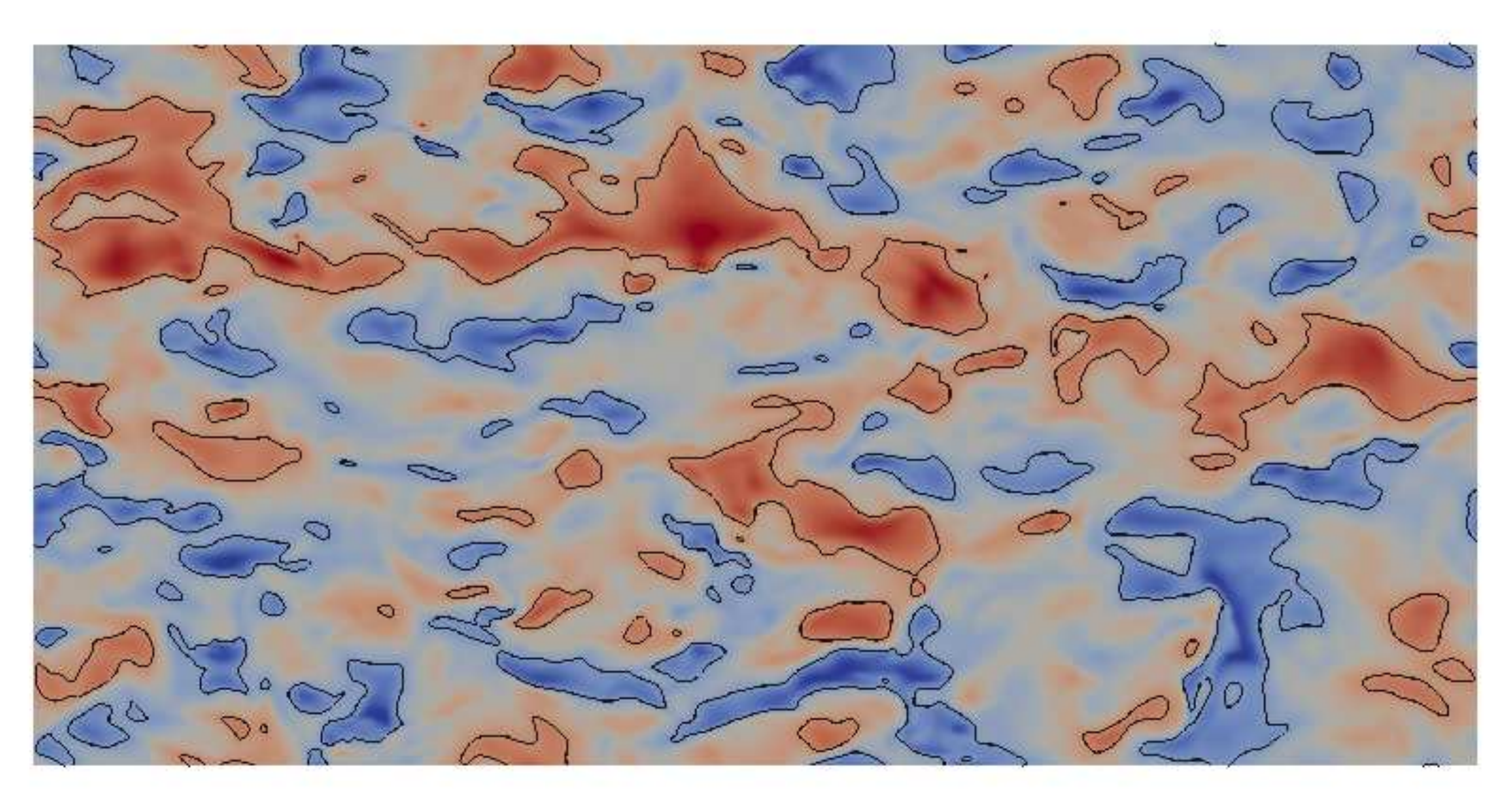}
  \caption{Instantaneous contour of the streamwise velocity fluctuations $u'$ on the $x-z$ plane at $\left( \tilde{y}+d \right)^+=10$ for walls with transverse elastic modulus $G$ decreasing from top-left to bottom-right. The color scale goes from $-0.35U_b$ (blue) to $0.35U_b$ (red). The black lines represent the levels $u'^+ \pm 4$.}
  \label{fig:sliceXZ-Ufluct-yp}
\end{figure}

\begin{figure}
  \centering
  \includegraphics[width=0.45\textwidth]{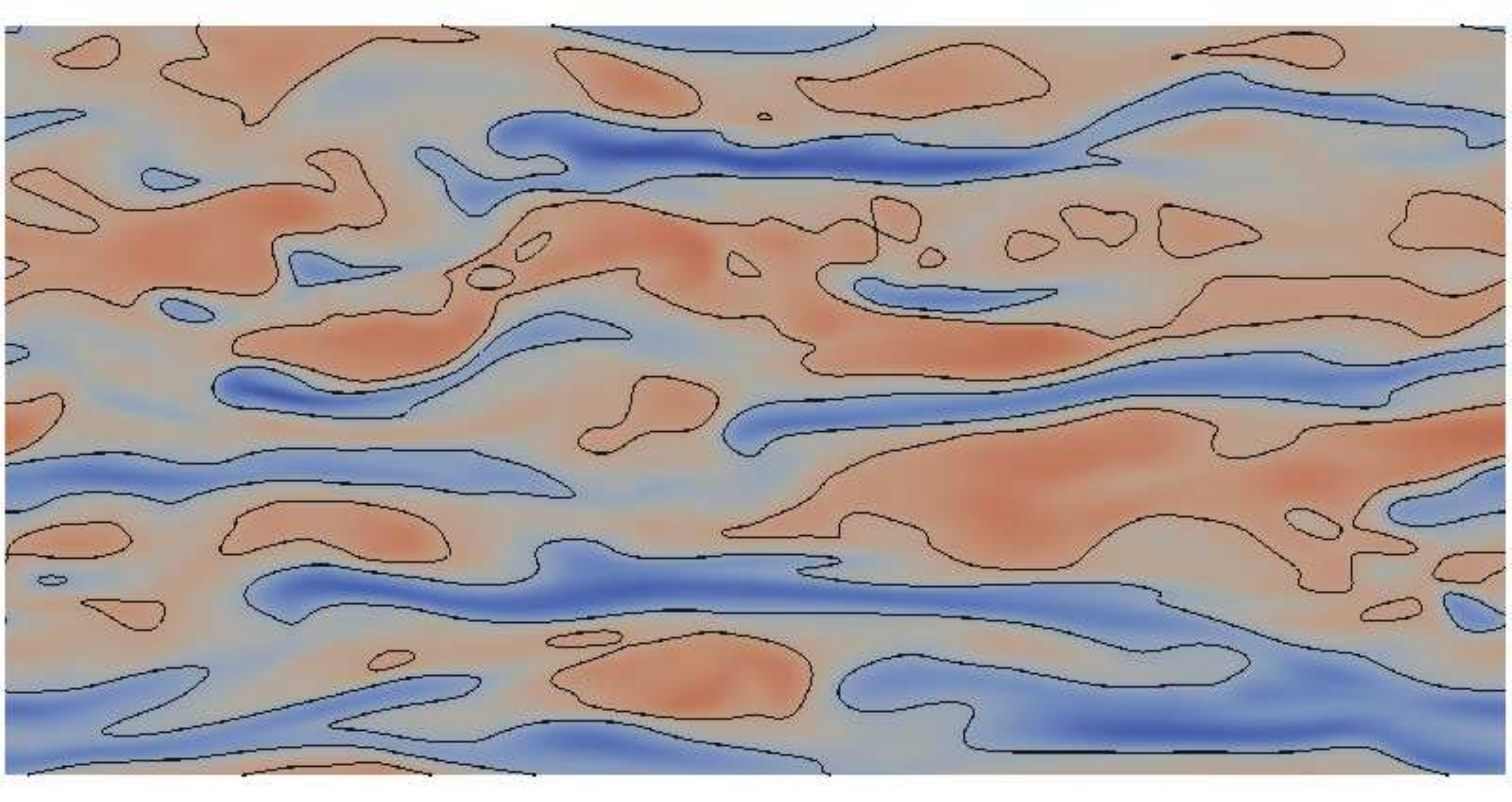}
  \includegraphics[width=0.45\textwidth]{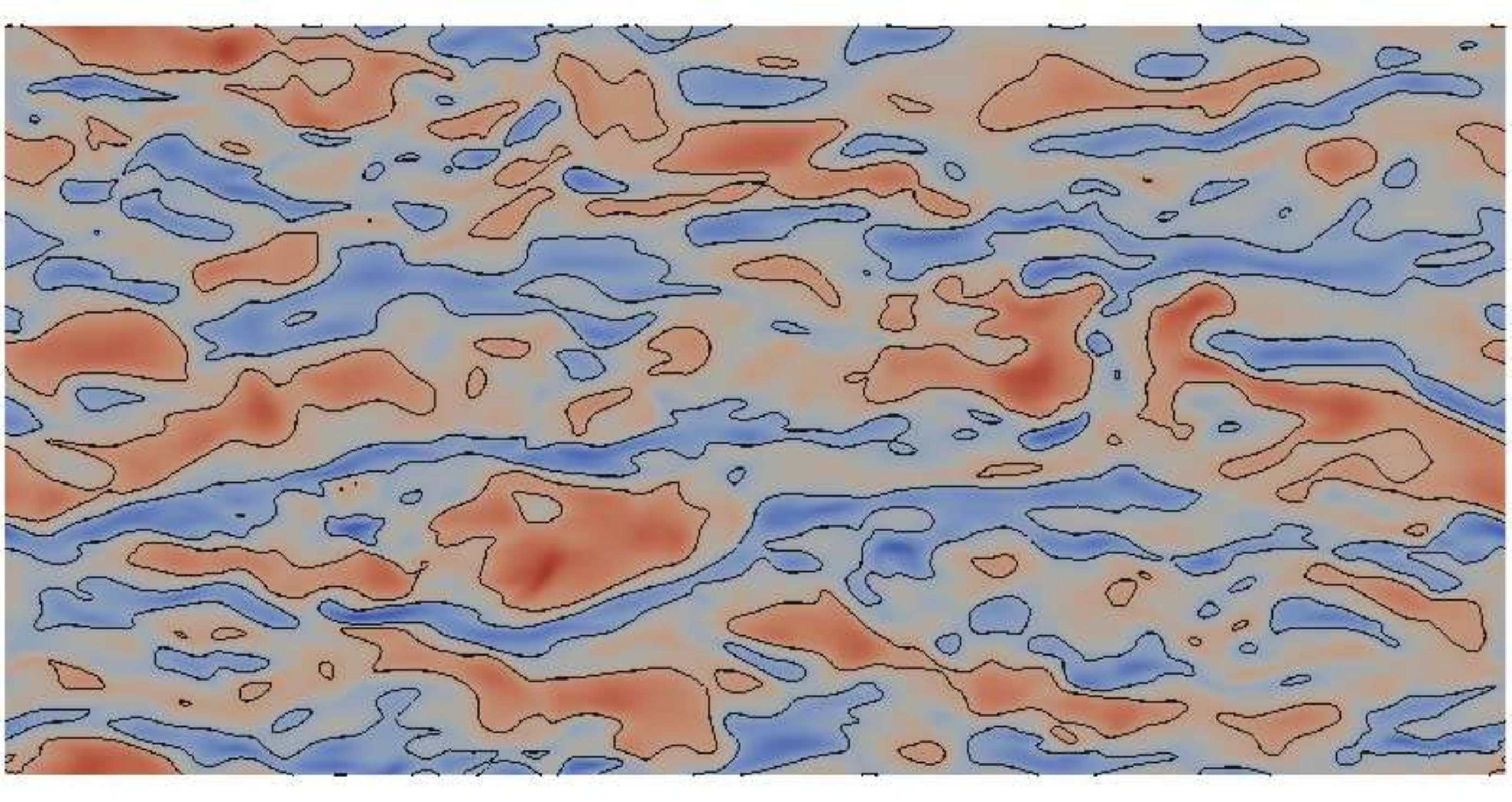} \\ \vspace{0.8mm}
  \includegraphics[width=0.45\textwidth]{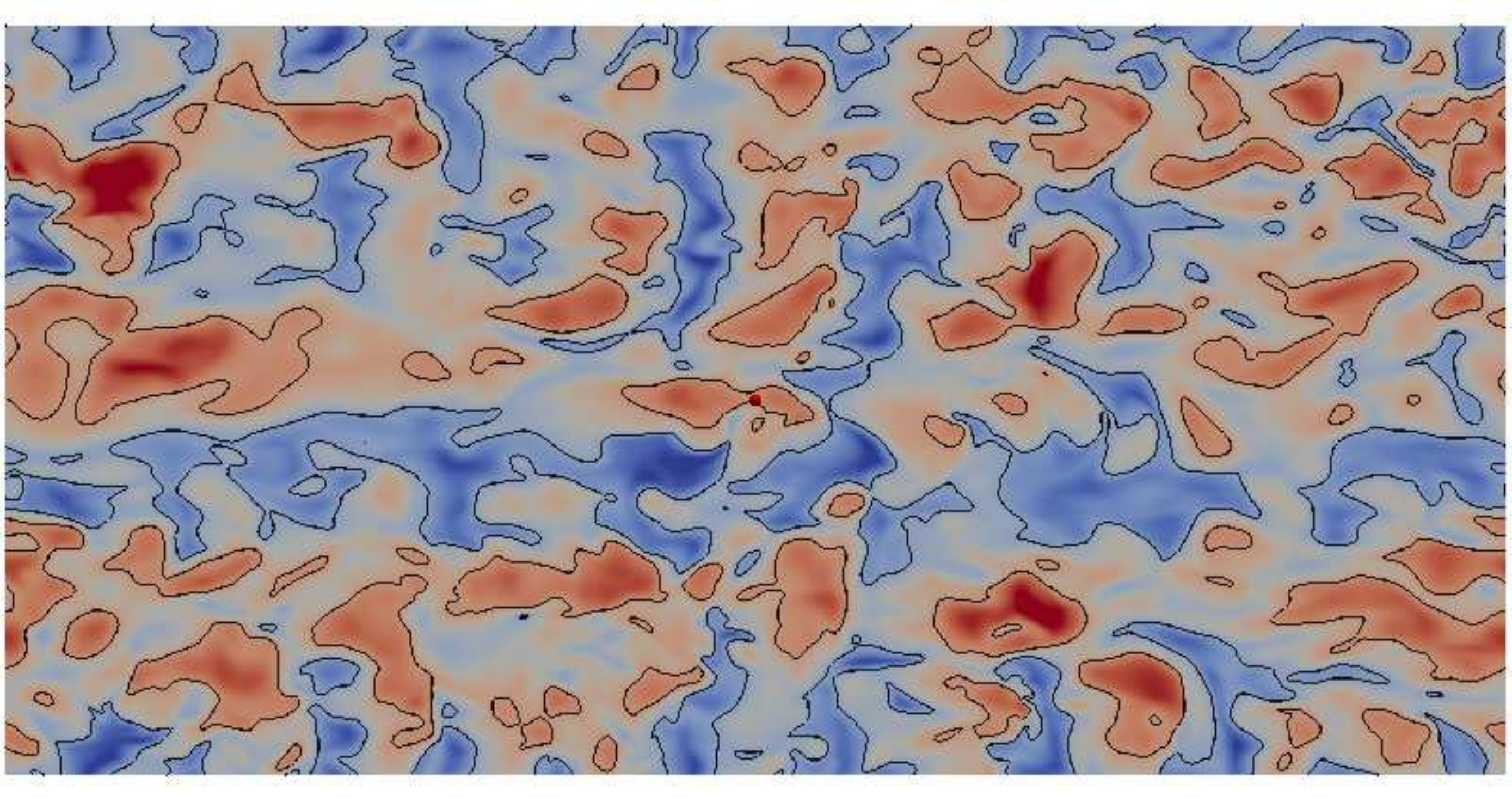}
  \includegraphics[width=0.45\textwidth]{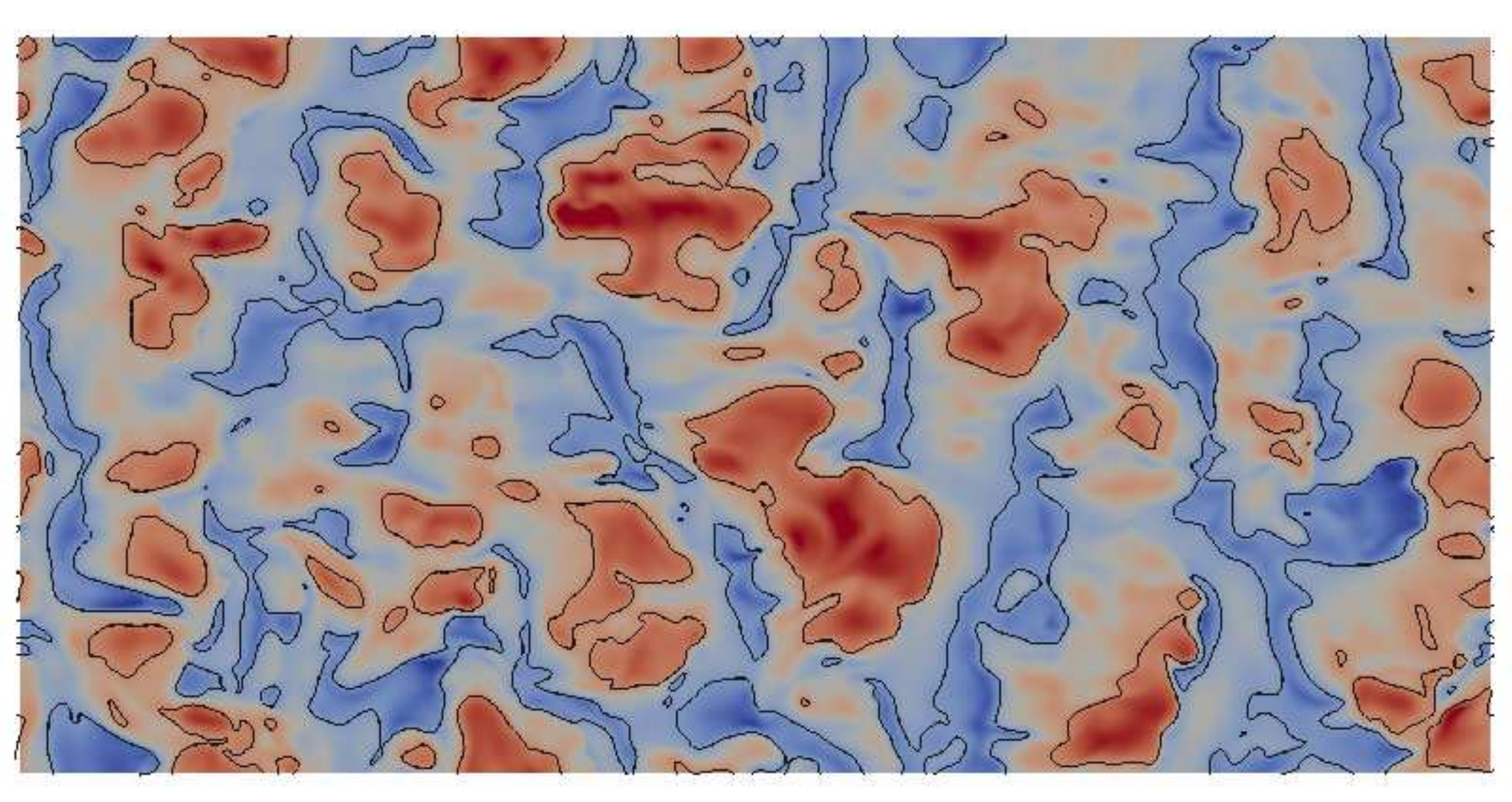}
  \caption{Instantaneous contour of the streamwise velocity fluctuations $u'$ on the wall-parallel $x-z$ plane at $\tilde{y}=0.15h$ for walls with transverse elastic modulus $G$ decreasing from top-left to bottom-right. The color scale goes from $-0.35U_b$ (blue) to $0.35U_b$ (red). The black lines represent the levels $u'^+ \pm 4$.}
  \label{fig:sliceXZ-Ufluct-y}
\end{figure}

The wall deformation affects the near-wall structure of the flow and this is visually confirmed by \figrefA{fig:sliceXZ-Ufluct-yp} that depicts the flow  in the wall-parallel  $x-z$ plane at $\left( \tilde{y}+d \right)^+=10$. The figure reports instantaneous contours of the streamwise velocity fluctuations $u'$ with color scale from $-0.35U_b$ (blue) to $0.35U_b$ (red). In the same figure, we also show the levels $u'^+ = \pm 4$ using black line contours. The goal of this figure is to identify the large-scale low- and high-speed near-wall streaks and show how they are affected by the wall elasticity. It is evident that the structures appear less elongated and more fragmented. Indeed, the flow is rich of small-scale features, consistent with a picture where the larger coherent structures are being broken into smaller pieces as a result of the wall movement. \figrefAC{fig:sliceXZ-Ufluct-y} reports the same quantity but further from the wall, over the wall-parallel plane at $\tilde{y}=0.15h$. As for the near-wall structures the low- and high-speed streaks become shorter the more elastic the walls. Another interesting flow feature can be inferred by this figure, \ie as the structure become shorter they also increase their spanwise size, eventually extending almost over the entire spanwise extent of our domain (see \figref[d]{fig:sliceXZ-Ufluct-y}), consistently with our previous discussion about the wall deformation (see \figref[d]{fig:moving-wall}). These visual observations will be now quantified statistically by analysing the autocorrelation functions.

\begin{figure}
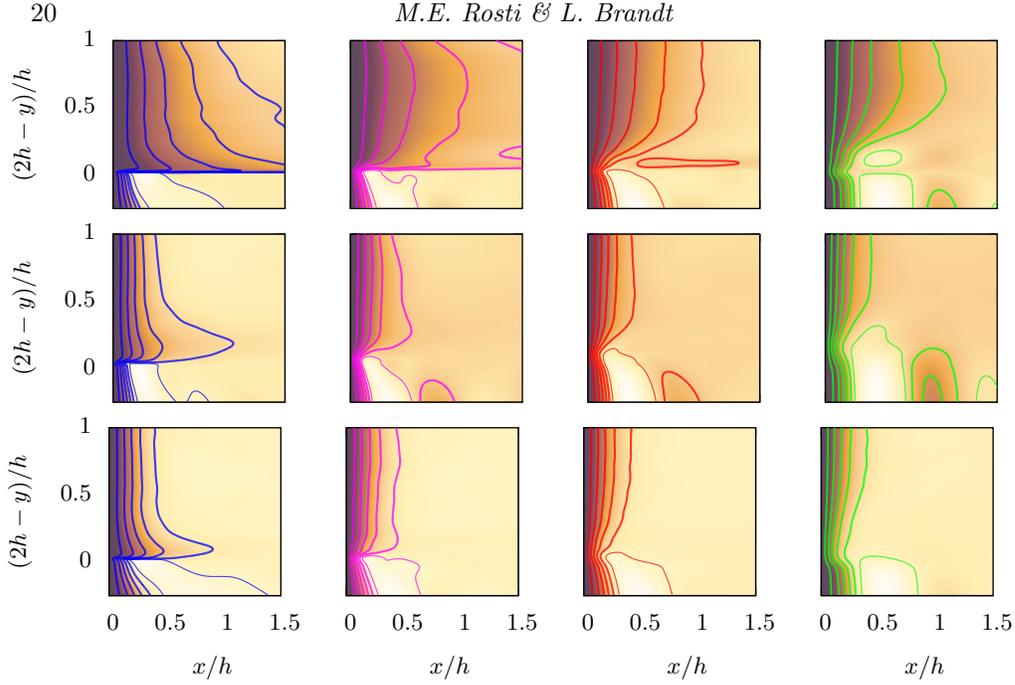

  \centering
  \input{fig14a}
  \input{fig14b}
  \input{fig14c}
  \input{fig14d} \\
  \input{fig14e}
  \input{fig14f}
  \input{fig14g}
  \input{fig14h} \\
  \input{fig14i}
  \input{fig14j}
  \input{fig14k}
  \input{fig14l}
  \vspace{0.8cm}
  \caption{Line and colour contours of the one-dimensional autocorrelation of the (top) streamwise, (middle) wall-normal, and (bottom) spanwise velocity fluctuations as a function of the streamwise spacing $x/h$ for different $y$ coordinates, shown across half elastic wall and half channel. The thick and thin lines correspond to positive and negative values, ranging from $-0.1$ to $0.9$ with a step of $0.2$ between two neighbouring lines. The color scale ranges from $-0.1$ (white) to $1.0$ (black).}
  \label{fig:corr-x}
\end{figure}

We now focus on the two-point correlation, defined as follows
\begin{equation} \label{eq:correlation}
R_{ii}(\bm{x},\bm{r})=\frac{\overline{u_i'(\bm{x})u_i'(\bm{x}+\bm{r})}}{\overline{u_i'^2(\bm{x})}},
\end{equation}
where the bar denotes average over time and homogeneous directions, and the prime the velocity fluctuation. \figrefAC{fig:corr-x} illustrates the structure of the  correlation functions along the streamwise direction $x$. These are one-dimensional correlations using the velocity difference along $x$, displayed for different $y$ coordinates across half of the elastic wall and the channel, \ie for $y\in\left[ 1h,2.25h \right]$. For the most rigid wall, the correlations, and in particular that for the streamwise component $u$, present in the fluid region structures of relatively large length especially close to the walls. These are associated to the presence of elongated streaky structures that characterize the near-wall region. For a completely rigid wall, the streaks have typically a length of the order of $1000$ wall units, which would correspond to a correlation distance of roughly $5h$. As already seen in the visualisation in \figrefA{fig:sliceXZ-Ufluct-yp}, the typical lengthscale is strongly reduced when increasing the wall elasticity, and indeed the correlation distance decreases significantly close to the wall, which is consistent with the reduction/absence of low- and high-speed streaks.

It is also interesting to note that the correlations change rapidly but continuously across the interface: the typical correlation length becomes much shorter for the $u$ component inside the elastic material as this cannot support  streaky structures, while it remains comparable to the rigid-channel lengthscales for the other two velocity components. Inside the elastic wall, the correlations present an alternating sequence of progressively weaker negative and positive local peaks, a so-called cell-like pattern, associated to large-scale pressure fluctuations within the elastic material and just above the interface, as suggested by \citet{breugem_boersma_uittenbogaard_2006a} for the flow over a porous wall. The typical wavelength of this oscillation for the most deformable case is approximately $1h$, thus corresponding to around $300$ plus units, a result consistent with the undulations shown in \figref[d]{fig:sliceXZ-Ufluct-yp}.

\begin{figure}
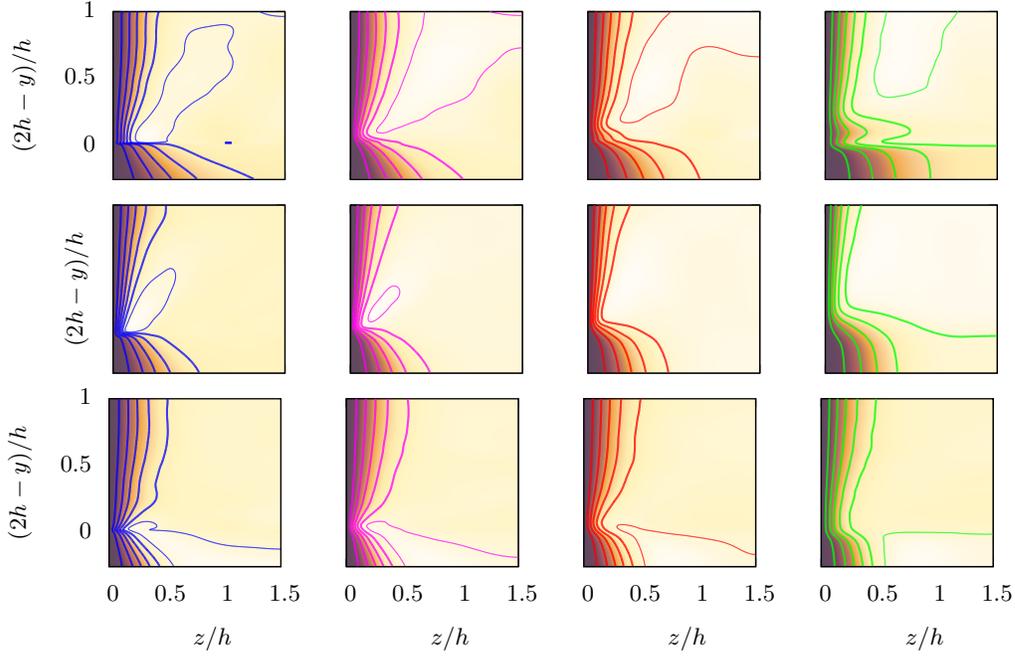

  \centering
  \input{fig15a}
  \input{fig15b}
  \input{fig15c}
  \input{fig15d} \\
  \input{fig15e}
  \input{fig15f}
  \input{fig15g}
  \input{fig15h} \\
  \input{fig15i}
  \input{fig15j}
  \input{fig15k}
  \input{fig15l}
  \vspace{0.8cm}
  \caption{Line and colour contours of the one-dimensional autocorrelation of the (top) streamwise, (middle) wall-normal, and (bottom) spanwise velocity fluctuations as a function of the spanwise spacing $z/h$ for different $y$ coordinates, shown across half elastic wall and half channel. The thick and thin lines correspond to positive and negative values, ranging from $-0.1$ to $0.9$, with a step of $0.2$ between two neighbouring lines. The color scale ranges from $-0.1$ (white) to $1.0$ (black).}
  \label{fig:corr-z}
\end{figure}

Finally, we consider the spanwise correlation functions for all three velocity components, see \figrefA{fig:corr-z}. On a rigid wall \citep{kim_moin_moser_1987a}, the spanwise autocorrelation of the streamwise velocity component $u$ (\figref[top]{fig:corr-z}) exhibits a local minimum at $z^{w+}\approx 50$ in the region close to the wall, usually associated with the average spanwise distance between a low-speed and a neighbouring high-speed streak. Note that the periodicity of the streaks is deduced from the oscillations of the autocorrelation for larger spanwise spacings. The flow over elastic walls, conversely, display a larger correlation distance. Considering the wall-normal component of the velocity (\figref[middle]{fig:corr-z}), we observe a local minimum at roughly $y^{w+} \approx 25$ close to the rigid wall, which is consistent with the presence of quasi-streamwise vortices. The value of this local minimum decreases ($G\downarrow$ and $G\downarrow\downarrow$), and eventually disappear ($G\downarrow\downarrow\downarrow$ and $G\downarrow\downarrow\downarrow\downarrow$) in the case of moving elastic walls, suggesting that the quasi-streamwise vortices are strongly reduced. All the three velocity components present relatively large spanwise scales, and we thus find a statistical signature of the spanwise vortical structures or rolls similarly to those that have been identified in the turbulent flow over porous walls \citep{breugem_boersma_uittenbogaard_2006a, rosti_cortelezzi_quadrio_2015a, samanta_vinuesa_lashgari_schlatter_brandt_2015a}, rough surfaces \citep{jimenez_uhlmann_pinelli_kawahara_2001a} and plant canopies \citep{finnigan_2000a}. These structures are also believed  to be the main cause of the performance loss of drag-reducing riblets when the riblets size is increased above their optimum \citep{garcia-mayoral_jimenez_2011a}.

\subsection{Effect of viscosity ratio} \label{sec:result-solidvisc}
\begin{figure}
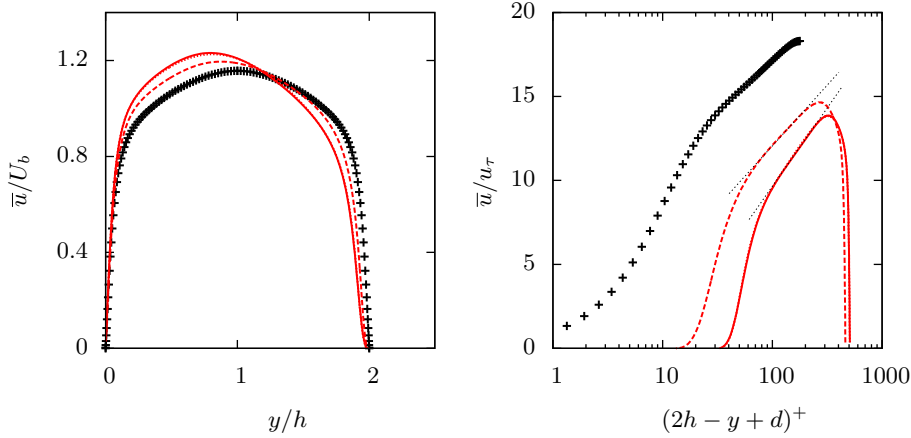

  \centering
  \input{fig16a}
  \input{fig16b}
  \vspace{0.5cm}
  \caption{(a) Comparison of the streamwise mean velocity profile $U$ of a turbulent channel flow at $Re=2800$ over rigid walls ($+$ symbols) and over different hyper-elastic walls (red lines). The dotted, solid, and dashed lines are used for the cases $\mu^s/\mu^f=0.1$, $1.0$, and $10.0$ at a fixed elastic modulus $G=1.0$. The results for the rigid case are taken from \citet{kim_moin_moser_1987a}. (b) Mean velocity profile $U$ versus the distance from the top deformable wall in wall units.}
  \label{fig:visc-mean-vel}
\end{figure}

We now consider the effect of the viscosity of the hyper-elastic material on the turbulence statistics for the case $G\downarrow\downarrow\downarrow$. Three viscosity ratios are examined, \ie $\mu^s/\mu^f=0.1$, $1.0$, and $10.0$, see \tabref{tab:cases} for a direct comparison of the values of the main global statistics. 

When the elastic layer is less viscous than the fluid ($\mu^s\downarrow$), the friction Reynolds number is equal to $240$, same value obtained when the two viscosities are the same, while when the layer is more viscous than the fluid ($\mu^s\uparrow$), the Reynolds number decreases by $6\%$ to $226$. This trend is shown by the mean velocity  profiles, reported in \figref[a]{fig:visc-mean-vel}, where the dotted, solid, and dashed lines correspond to the flow with $\mu^s/\mu^f=0.1$, $1.0$, and $10.0$, respectively. 

\begin{figure}
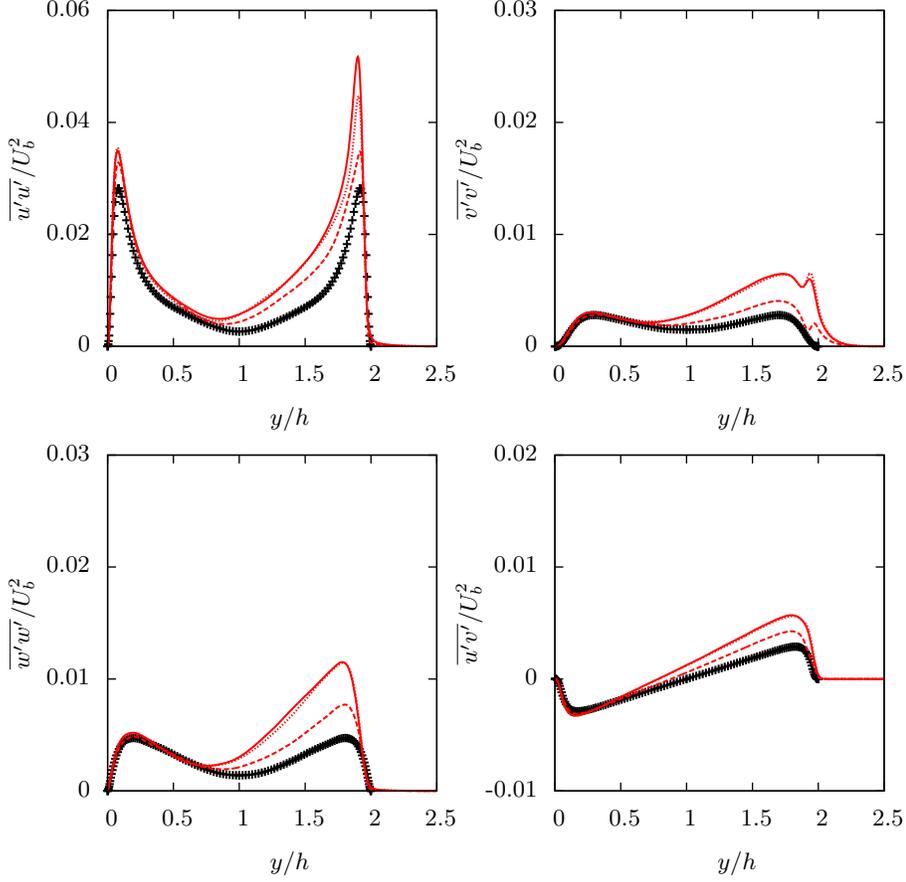

  \centering
  \input{fig17a}
  \input{fig17b}
  \vspace{0.8cm} \\
  \input{fig17c}
  \input{fig17d}
  \vspace{0.5cm}
  \caption{Components of the Reynolds stress tensor as a function of the wall-normal coordinate $y$, normalised with the bulk velocity $U_b$ for a turbulent channel flow over a hyper-elastic wall. Symbols are for the rigid case from \citet{kim_moin_moser_1987a} while the dotted, solid, and dashed lines are used for the cases $\mu^s/\mu^f=0.1$, $1.0$, and $10.0$ at a fixed elastic modulus $G=1.0$. The first three panels (a), (b), and (c) show the diagonal components $u'u'$, $v'v'$, and $w'w'$, whereas the cross term $u'v'$ is depicted in  (d).}
  \label{fig:visc-rms}
\end{figure}

The mean velocity profile is less skewed in the high viscosity case ($\mu^s\uparrow$), with its maximum velocity $u^M$ decreasing and located closer to the centreline. The same velocity profiles are displayed versus the logarithm of the distance from the top wall in wall-units in \figref[b]{fig:visc-mean-vel}. Again, the cases with $\mu^s<\mu^f$ and $\mu^s=\mu^f$ are almost undistinguishable, while  the downward shift of the profile is reduced when $\mu^s<\mu^f$, moving the logarithmic part of the velocity profile towards that for a rigid wall. A similar picture emerges after examining other quantities, such the wall-normal profile of the Reynolds stress tensor components reported in figure \ref{fig:visc-rms}.

Overall, the results suggest that the behaviour of the flow in the limit of large and small values of $\mu^s$ and $G$ is different. Indeed, as $G$ or $\mu^s/\mu^f$ tend to infinity, the wall deformation decreases and the flow approaches the one over a rigid wall. On the contrary, while $G$ tending to $0$ increases the wall deformation, thus enhancing the turbulence, the flow appears to be almost unaffected by a decrease of $\mu^s/\mu^f$ below unity. Indeed, this behaviour suggests that when the viscosity ratio is low, the deformation is determined by the fluid fluctuations, while when the ratio $\mu^s/\mu^f$ is high the deformation is limited by the properties of the elastic layer.

\section{Conclusion} \label{sec:conclusion}
We have carried out a number of direct numerical simulations of turbulent channel flow over a viscous hyper-elastic wall. The flow inside the fluid region is described by the Navier-Stokes equations, while momentum conservation and incompressibility are imposed inside the solid layer. The two sets of equations are coupled using a one-continuum formulation allowing a fully Eulerian description of the multiphase flow problem. We collect statistics by varying the values of the parameters involved in the definition of the elastic material (elasticity and viscosity) to assess the sensitivity of the flow statistics to each of them. 

The data show that the elasticity $G$ emerges as the key parameter. The turbulent flow in the channel is affected by the deformable wall even at low values of elasticity, where non-zero fluctuations of the vertical velocity at the interface influence the flow dynamics. In particular, the mean friction coefficient strongly increases with elasticity. Notwithstanding the zero mean value of the velocity of the elastic wall, the mean profile is modified by a downward shift of the inertial range and by an increase of its slope. The value of the wall-normal velocity fluctuation well correlates with the downward shift of the mean velocity profile and with the change of slope in the logarithmic region. This result appears to be valid also for the mean velocity profile of turbulent channel flows over highly permeable walls, and extends therefore the correlation proposed by \citet{orlandi_leonardi_2008a}  for turbulent channel flows over rough walls.

The penetration depth of the turbulent motions within the solid layer seems to primarily depend upon $G$. A secondary peak in the wall-normal profile of the velocity fluctuations associated with the movement of the wall appears at the interface, which compensates for the decrease of the streamwise velocity fluctuations in wall units, so  that the turbulent kinetic energy remains almost constant. This secondary peak is located around the maximum wall oscillation and is generated by the wall motion. The decrease in the streamwise fluctuations is associated to the disruption of the quasi streamwise vortices and low- and high-speed streaks near the deformable wall, as confirmed by the instantaneous visualisations and one-dimensional velocity correlations. These also show the increase of the flow coherence in the spanwise direction for increasing elasticities, as observed for flows over porous walls.

Finally, we examine the role of the viscosity of the elastic layer: the data reveal that changes in the turbulence statistics appear for materials which are more viscous than the fluid, with an overall reduction of the friction Reynolds number and of the velocity fluctuations, with values approaching those for a rigid wall. No significant changes are found for low viscosity ratios, $\mu^s/\mu^f$, indicating that when the viscosity of the elastic layer is low, the deformation is determined by the fluid fluctuations, while when the viscosity ratio is high the deformation is limited by the solid properties.

The analysis performed here assumes possibly the simplest behavior for the  hyper-elastic solid. The present results can therefore be extended in a number of non-trivial ways, not only considering a more complex constitutive equation, \eg anisotropic and inhomogeneous material as well as poroelastic substrates, but also more complex flow configurations, \eg separating flows over elastic walls.

\section*{Acknowledgment}
This work was supported by the European Research Council Grant no. ERC-2013-CoG-616186, TRITOS and by the Swedish Research Council Grant no. VR 2014-5001. The authors acknowledge computer time provided by SNIC (Swedish National Infrastructure for Computing).

\bibliographystyle{jfm}
\bibliography{bibliography.bib}

\begin{thebibliography}{71}
\expandafter\ifx\csname natexlab\endcsname\relax\def\natexlab#1{#1}\fi

\bibitem[Antonia \& Krogstad(2001)]{antonia_krogstad_2001a}
{\sc Antonia, R~A \& Krogstad, P~A} 2001 Turbulence structure in boundary
  layers over different types of surface roughness. {\em {F}luid {D}ynamics
  {R}esearch\/} {\bf 28}~(2), 139--157.

\bibitem[Beavers {\em et~al.\/}(1970)Beavers, Sparrow \&
  Magnuson]{beavers_sparrow_magnuson_1970a}
{\sc Beavers, G~S, Sparrow, E~M \& Magnuson, R~A} 1970 Experiments on coupled
  parallel flows in a channel and a bounding porous medium. {\em {J}ournal of
  {B}asic {E}ngineering\/} {\bf 92}, 843--848.

\bibitem[Belcher {\em et~al.\/}(2003)Belcher, Jerram \&
  Hunt]{belcher_jerram_hunt_2003a}
{\sc Belcher, S~E, Jerram, N \& Hunt, J C~R} 2003 Adjustment of a turbulent
  boundary layer to a canopy of roughness elements. {\em {J}ournal of {F}luid
  {M}echanics\/} {\bf 488}, 369--398.

\bibitem[Benjamin(1960)]{benjamin_1960a}
{\sc Benjamin, T~B} 1960 Effects of a flexible boundary on hydrodynamic
  stability. {\em {J}ournal of {F}luid {M}echanics\/} {\bf 9}~(04), 513--532.

\bibitem[Bonet \& Wood(1997)]{bonet_wood_1997a}
{\sc Bonet, J \& Wood, R~D} 1997 {\em Nonlinear continuum mechanics for finite
  element analysis\/}. {C}ambridge {U}niversity {P}ress.

\bibitem[Breugem {\em et~al.\/}(2006)Breugem, Boersma \&
  Uittenbogaard]{breugem_boersma_uittenbogaard_2006a}
{\sc Breugem, W~P, Boersma, B~J \& Uittenbogaard, R~E} 2006 The influence of
  wall permeability on turbulent channel flow. {\em {J}ournal of {F}luid
  {M}echanics\/} {\bf 562}, 35--72.

\bibitem[Bushnell {\em et~al.\/}(1977)Bushnell, Hefner \&
  Ash]{bushnell_hefner_ash_1977a}
{\sc Bushnell, D~M, Hefner, J~N \& Ash, R~L} 1977 Effect of compliant wall
  motion on turbulent boundary layers. {\em {P}hysics of {F}luids {A}: {F}luid
  {D}ynamics (1989-1993)\/} {\bf 20}~(10), S31--S48.

\bibitem[Cabal {\em et~al.\/}(2002)Cabal, Szumbarski \&
  Floryan]{cabal_szumbarski_floryan_2002a}
{\sc Cabal, A, Szumbarski, J \& Floryan, J~M} 2002 Stability of flow in a wavy
  channel. {\em {J}ournal of {F}luid {M}echanics\/} {\bf 457}, 191--212.

\bibitem[Carpenter \& Garrad(1985)]{carpenter_garrad_1985a}
{\sc Carpenter, P~W \& Garrad, A~D} 1985 The hydrodynamic stability of flow
  over {K}ramer-type compliant surfaces. {P}art 1. {T}ollmien-{S}chlichting
  instabilities. {\em {J}ournal of {F}luid {M}echanics\/} {\bf 155}, 465--510.

\bibitem[Carpenter \& Morris(1990)]{carpenter_morris_1990a}
{\sc Carpenter, P~W \& Morris, P~J} 1990 The effect of anisotropic wall
  compliance on boundary-layer stability and transition. {\em {J}ournal of
  {F}luid {M}echanics\/} {\bf 218}, 171--223.

\bibitem[Chang {\em et~al.\/}(1996)Chang, Hou, Merriman \&
  Osher]{chang_hou_merriman_osher_1996a}
{\sc Chang, Y~C, Hou, T~Y, Merriman, B \& Osher, S} 1996 A level set
  formulation of eulerian interface capturing methods for incompressible fluid
  flows. {\em {J}ournal of {C}omputational {P}hysics\/} {\bf 124}~(2),
  449--464.

\bibitem[Cheng \& Castro(2002)]{cheng_castro_2002a}
{\sc Cheng, H \& Castro, I~P} 2002 Near wall flow over urban-like roughness.
  {\em {B}oundary-{L}ayer {M}eteorology\/} {\bf 104}~(2), 229--259.

\bibitem[Choi {\em et~al.\/}(1997)Choi, Yang, Clayton, Glover, Atlar, Semenov
  \& Kulik]{choi_yang_clayton_glover_atlar_semenov_kulik_1997a}
{\sc Choi, K~S, Yang, X, Clayton, B~R, Glover, E~J, Atlar, M, Semenov, B~N \&
  Kulik, V~M} 1997 Turbulent drag reduction using compliant surfaces. In {\em
  {P}roceedings of the {R}oyal {S}ociety of {L}ondon {A}: {M}athematical,
  {P}hysical and {E}ngineering {S}ciences\/}, , vol. 453, pp. 2229--2240. {T}he
  {R}oyal {S}ociety.

\bibitem[Clauser(1954)]{clauser_1954a}
{\sc Clauser, F~H} 1954 Turbulent boundary layers in adverse pressure
  gradients. {\em {J}ournal of {A}erosol {S}cience\/} {\bf 21}, 91--109.

\bibitem[Daniel {\em et~al.\/}(1987)Daniel, Gaster \&
  Willis]{daniel_gaster_willis_1987a}
{\sc Daniel, A~P, Gaster, M \& Willis, G J~K} 1987 Boundary layer stability on
  compliant surfaces. {\em Tech. Rep.\/} 35020. {B}ritish {M}aritime
  {T}echnology {L}td.

\bibitem[Davies \& Carpenter(1997)]{davies_carpenter_1997a}
{\sc Davies, Christopher \& Carpenter, Peter~W} 1997 Numerical simulation of
  the evolution of {T}ollmien--{S}chlichting waves over finite compliant
  panels. {\em {J}ournal of {F}luid {M}echanics\/} {\bf 335}, 361--392.

\bibitem[Finnigan(2000)]{finnigan_2000a}
{\sc Finnigan, J} 2000 Turbulence in plant canopies. {\em {A}nnual {R}eview of
  {F}luid {M}echanics\/} {\bf 32}~(1), 519--571.

\bibitem[Flores \& Jimenez(2006)]{flores_jimenez_2006a}
{\sc Flores, O \& Jimenez, J} 2006 Effect of wall-boundary disturbances on
  turbulent channel flows. {\em {J}ournal of {F}luid {M}echanics\/} {\bf 566},
  357--376.

\bibitem[Garcia-Mayoral \& Jimenez(2011)]{garcia-mayoral_jimenez_2011a}
{\sc Garcia-Mayoral, R \& Jimenez, J} 2011 Drag reduction by riblets. {\em
  {P}hilosophical {T}ransactions of the {R}oyal {S}ociety of {L}ondon {A}:
  {M}athematical, {P}hysical and {E}ngineering {S}ciences\/} {\bf 369}~(1940),
  1412--1427.

\bibitem[Gaster(1988)]{gaster_1988a}
{\sc Gaster, M} 1988 Is the dolphin a red herring? In {\em Turbulence
  management and relaminarisation\/}, pp. 285--304. {S}pringer.

\bibitem[Gad-el Hak(1986)]{gad-el-hak_1986a}
{\sc Gad-el Hak, M} 1986 The response of elastic and viscoelastic surfaces to a
  turbulent boundary layer. {\em {J}ournal of {A}pplied {M}echanics\/} {\bf
  53}~(1), 206--212.

\bibitem[Gad-el Hak(1987)]{gad-el-hak_1987a}
{\sc Gad-el Hak, M} 1987 Compliant coatings research: a guide to the
  experimentalist. {\em {J}ournal of {F}luids and {S}tructures\/} {\bf 1}~(1),
  55--70.

\bibitem[Gad-el Hak {\em et~al.\/}(1996)]{gad-el-hak_others_1996a}
{\sc Gad-el Hak, Mohamed {\em et~al.\/}} 1996 Compliant coatings: a decade of
  progress. {\em {A}pplied {M}echanics {R}eviews\/} {\bf 49}, S147--S160.

\bibitem[Hama(1954)]{hama_1954a}
{\sc Hama, F~R} 1954 {\em Boundary-layer characteristics for smooth and rough
  surfaces\/}. {SNAME}.

\bibitem[Hirt \& Nichols(1981)]{hirt_nichols_1981a}
{\sc Hirt, C~W \& Nichols, B~D} 1981 Volume of fluid ({VOF}) method for the
  dynamics of free boundaries. {\em {J}ournal of {C}omputational {P}hysics\/}
  {\bf 39}~(1), 201--225.

\bibitem[Jackson(1981)]{jackson_1981a}
{\sc Jackson, P~S} 1981 On the displacement height in the logarithmic velocity
  profile. {\em {J}ournal of {F}luid {M}echanics\/} {\bf 111}, 15--25.

\bibitem[Jimenez {\em et~al.\/}(2001)Jimenez, Uhlmann, Pinelli \&
  Kawahara]{jimenez_uhlmann_pinelli_kawahara_2001a}
{\sc Jimenez, J, Uhlmann, M, Pinelli, A \& Kawahara, G} 2001 Turbulent shear
  flow over active and passive porous surfaces. {\em {J}ournal of {F}luid
  {M}echanics\/} {\bf 442}, 89--117.

\bibitem[Kim \& Moin(1985)]{kim_moin_1985a}
{\sc Kim, J \& Moin, P} 1985 Application of a fractional-step method to
  incompressible navier-stokes equations. {\em {J}ournal of {C}omputational
  {P}hysics\/} {\bf 59}~(2), 308--323.

\bibitem[Kim {\em et~al.\/}(1987)Kim, Moin \& Moser]{kim_moin_moser_1987a}
{\sc Kim, J, Moin, P \& Moser, R} 1987 Turbulence statistics in fully developed
  channel flow at low {R}eynolds number. {\em {J}ournal of {F}luid
  {M}echanics\/} {\bf 177}, 133--166.

\bibitem[Krindel \& Silberberg(1979)]{krindel_silberberg_1979a}
{\sc Krindel, P \& Silberberg, A} 1979 Flow through gel-walled tubes. {\em
  {J}ournal of {C}olloid and {I}nterface {S}cience\/} {\bf 71}~(1), 39--50.

\bibitem[Krogstad {\em et~al.\/}(1992)Krogstad, Antonia \&
  Browne]{krogstad_antonia_browne_1992a}
{\sc Krogstad, P~A, Antonia, R~A \& Browne, L W~B} 1992 Comparison between
  rough-and smooth-wall turbulent boundary layers. {\em {J}ournal of {F}luid
  {M}echanics\/} {\bf 245}, 599--617.

\bibitem[Krogstadt \& Antonia(1999)]{krogstadt_antonia_1999a}
{\sc Krogstadt, P~A \& Antonia, R~A} 1999 Surface roughness effects in
  turbulent boundary layers. {\em {E}xperiments in {F}luids\/} {\bf 27}~(5),
  450--460.

\bibitem[Kumaran(1995)]{kumaran_1995a}
{\sc Kumaran, V} 1995 Stability of the flow of a fluid through a flexible tube
  at high reynolds number. {\em {J}ournal of {F}luid {M}echanics\/} {\bf 302},
  117--139.

\bibitem[Kumaran(1996)]{kumaran_1996a}
{\sc Kumaran, V} 1996 Stability of inviscid flow in a flexible tube. {\em
  {J}ournal of {F}luid {M}echanics\/} {\bf 320}, 1--17.

\bibitem[Kumaran(1998{\natexlab{{\em a\/}}})]{kumaran_1998a}
{\sc Kumaran, V} 1998{\natexlab{{\em a\/}}} Stability of the flow of a fluid
  through a flexible tube at intermediate reynolds number. {\em {J}ournal of
  {F}luid {M}echanics\/} {\bf 357}, 123--140.

\bibitem[Kumaran(1998{\natexlab{{\em b\/}}})]{kumaran_1998b}
{\sc Kumaran, V} 1998{\natexlab{{\em b\/}}} Stability of wall modes in a
  flexible tube. {\em {J}ournal of {F}luid {M}echanics\/} {\bf 362}, 1--15.

\bibitem[Kumaran {\em et~al.\/}(1994)Kumaran, Fredrickson \&
  Pincus]{kumaran_fredrickson_pincus_1994a}
{\sc Kumaran, V, Fredrickson, G~H \& Pincus, P} 1994 Flow induced instability
  of the interface between a fluid and a gel at low reynolds number. {\em
  {J}ournal de {P}hysique {II}\/} {\bf 4}~(6), 893--911.

\bibitem[Kumaran \& Muralikrishnan(2000)]{kumaran_muralikrishnan_2000a}
{\sc Kumaran, V \& Muralikrishnan, R} 2000 Spontaneous growth of fluctuations
  in the viscous flow of a fluid past a soft interface. {\em {P}hysical
  {R}eview {L}etters\/} {\bf 84}~(15), 3310.

\bibitem[Lahav {\em et~al.\/}(1973)Lahav, Eliezer \&
  Silberberg]{lahav_eliezer_silberberg_1973a}
{\sc Lahav, J, Eliezer, N \& Silberberg, A} 1973 Gel-walled cylindrical
  channels as models for the microcirculation: dynamics of flow. {\em
  {B}iorheology\/} {\bf 10}~(4), 595--604.

\bibitem[Landahl(1962)]{landahl_1962a}
{\sc Landahl, M~T} 1962 On the stability of a laminar incompressible boundary
  layer over a flexible surface. {\em {J}ournal of {F}luid {M}echanics\/} {\bf
  13}~(04), 609--632.

\bibitem[Lee {\em et~al.\/}(1993)Lee, Fisher \&
  Schwarz]{lee_fisher_schwarz_1993a}
{\sc Lee, T, Fisher, M \& Schwarz, WH} 1993 Investigation of the stable
  interaction of a passive compliant surface with a turbulent boundary layer.
  {\em {J}ournal of {F}luid {M}echanics\/} {\bf 257}, 373--401.

\bibitem[Leonardi {\em et~al.\/}(2007)Leonardi, Orlandi \&
  Antonia]{leonardi_orlandi_antonia_2007a}
{\sc Leonardi, S, Orlandi, P \& Antonia, R~A} 2007 Properties of d-and k-type
  roughness in a turbulent channel flow. {\em {P}hysics of {F}luids
  (1994-present)\/} {\bf 19}~(12), 125101.

\bibitem[Leonardi {\em et~al.\/}(2004)Leonardi, Orlandi, Djenidi \&
  Antonia]{leonardi_orlandi_djenidi_antonia_2004a}
{\sc Leonardi, S, Orlandi, P, Djenidi, L \& Antonia, R~A} 2004 Structure of
  turbulent channel flow with square bars on one wall. {\em {I}nternational
  {J}ournal of {H}eat and {F}luid {F}low\/} {\bf 25}~(3), 384--392.

\bibitem[Leonardi {\em et~al.\/}(2003)Leonardi, Orlandi, Smalley, Djenidi \&
  Antonia]{leonardi_orlandi_smalley_djenidi_antonia_2003a}
{\sc Leonardi, S, Orlandi, P, Smalley, R~J, Djenidi, L \& Antonia, R~A} 2003
  Direct numerical simulations of turbulent channel flow with transverse square
  bars on one wall. {\em {J}ournal of {F}luid {M}echanics\/} {\bf 491},
  229--238.

\bibitem[Luo \& Bewley(2003)]{luo_bewley_2003a}
{\sc Luo, H \& Bewley, T~R} 2003 Design, modeling, and optimization of
  compliant tensegrity fabrics for the reduction of turbulent skin friction. In
  {\em {S}mart {S}tructures and {M}aterials\/}, pp. 460--470. {I}nternational
  {S}ociety for {O}ptics and {P}hotonics.

\bibitem[Luo \& Bewley(2005)]{luo_bewley_2005a}
{\sc Luo, H \& Bewley, T~R} 2005 Accurate simulation of near-wall turbulence
  over a compliant tensegrity fabric. In {\em {S}mart {S}tructures and
  {M}aterials\/}, pp. 184--197. International Society for Optics and Photonics.

\bibitem[Min {\em et~al.\/}(2001)Min, Yoo \& Choi]{min_yoo_choi_2001a}
{\sc Min, T, Yoo, J~Y \& Choi, H} 2001 Effect of spatial discretization schemes
  on numerical solutions of viscoelastic fluid flows. {\em {J}ournal of
  {N}on-{N}ewtonian {F}luid {M}echanics\/} {\bf 100}~(1), 27--47.

\bibitem[Nikuradse(1933)]{nikuradse_1933a}
{\sc Nikuradse, J} 1933 Laws of flow in rough pipes. In {\em {VDI}
  {F}orschungsheft\/}. Citeseer.

\bibitem[Nikuradse(1950)]{nikuradse_1950a}
{\sc Nikuradse, J} 1950 Laws of flow in rough pipes. {\em Tech. Rep.\/}.
  {N}ational {A}dvisory {C}ommittee for {A}eronautics {W}ashington.

\bibitem[Orlandi \& Leonardi(2006)]{orlandi_leonardi_2006a}
{\sc Orlandi, P \& Leonardi, S} 2006 Dns of turbulent channel flows with
  two-and three-dimensional roughness. {\em {J}ournal of {T}urbulence\/} {\bf
  7}~(53), N73.

\bibitem[Orlandi \& Leonardi(2008)]{orlandi_leonardi_2008a}
{\sc Orlandi, P \& Leonardi, S} 2008 Direct numerical simulation of
  three-dimensional turbulent rough channels: parameterization and flow
  physics. {\em {J}ournal of {F}luid {M}echanics\/} {\bf 606}, 399--415.

\bibitem[Orlandi {\em et~al.\/}(2006)Orlandi, Leonardi \&
  Antonia]{orlandi_leonardi_antonia_2006a}
{\sc Orlandi, P, Leonardi, S \& Antonia, R~A} 2006 Turbulent channel flow with
  either transverse or longitudinal roughness elements on one wall. {\em
  {J}ournal of {F}luid {M}echanics\/} {\bf 561}, 279--305.

\bibitem[Orlandi {\em et~al.\/}(2003)Orlandi, Leonardi, Tuzi \&
  Antonia]{orlandi_leonardi_tuzi_antonia_2003a}
{\sc Orlandi, P, Leonardi, S, Tuzi, R \& Antonia, R~A} 2003 Direct numerical
  simulation of turbulent channel flow with wall velocity disturbances. {\em
  {P}hysics of {F}luids (1994-present)\/} {\bf 15}~(12), 3587--3601.

\bibitem[Perot \& Moin(1995{\natexlab{{\em a\/}}})]{perot_moin_1995a}
{\sc Perot, B \& Moin, P} 1995{\natexlab{{\em a\/}}} Shear-free turbulent
  boundary layers. {P}art 1. {P}hysical insights into near-wall turbulence.
  {\em {J}ournal of {F}luid {M}echanics\/} {\bf 295}, 199--227.

\bibitem[Perot \& Moin(1995{\natexlab{{\em b\/}}})]{perot_moin_1995b}
{\sc Perot, B \& Moin, P} 1995{\natexlab{{\em b\/}}} Shear-free turbulent
  boundary layers. {P}art 2. {N}ew concepts for {R}eynolds stress transport
  equation modelling of inhomogeneous flows. {\em {J}ournal of {F}luid
  {M}echanics\/} {\bf 295}, 229--245.

\bibitem[Picano {\em et~al.\/}(2015)Picano, Breugem \&
  Brandt]{picano_breugem_brandt_2015a}
{\sc Picano, F, Breugem, W~P \& Brandt, L} 2015 Turbulent channel flow of dense
  suspensions of neutrally buoyant spheres. {\em {J}ournal of {F}luid
  {M}echanics\/} {\bf 764}, 463--487.

\bibitem[Pluvinage {\em et~al.\/}(2014)Pluvinage, Kourta \&
  Bottaro]{pluvinage_kourta_bottaro_2014a}
{\sc Pluvinage, F, Kourta, A \& Bottaro, A} 2014 Instabilities in the boundary
  layer over a permeable, compliant wall. {\em {P}hysics of {F}luids
  (1994-present)\/} {\bf 26}~(8), 084103.

\bibitem[Quintard \& Whitaker(1994)]{quintard_whitaker_1994b}
{\sc Quintard, M \& Whitaker, S} 1994 Transport in ordered and disordered
  porous media ii: Generalized volume averaging. {\em {T}ransport in {P}orous
  {M}edia\/} {\bf 14}~(2), 179--206.

\bibitem[Rosti {\em et~al.\/}(2015)Rosti, Cortelezzi \&
  Quadrio]{rosti_cortelezzi_quadrio_2015a}
{\sc Rosti, M~E, Cortelezzi, L \& Quadrio, M} 2015 Direct numerical simulation
  of turbulent channel flow over porous walls. {\em {J}ournal of {F}luid
  {M}echanics\/} {\bf 784}, 396--442.

\bibitem[Samanta {\em et~al.\/}(2015)Samanta, Vinuesa, Lashgari, Schlatter \&
  Brandt]{samanta_vinuesa_lashgari_schlatter_brandt_2015a}
{\sc Samanta, A, Vinuesa, R, Lashgari, I, Schlatter, P \& Brandt, L} 2015
  Enhanced secondary motion of the turbulent flow through a porous square duct.
  {\em {J}ournal of {F}luid {M}echanics\/} {\bf 784}, 681--693.

\bibitem[Shankar \& Kumaran(1999)]{shankar_kumaran_1999a}
{\sc Shankar, V \& Kumaran, V} 1999 Stability of non-parabolic flow in a
  flexible tube. {\em {J}ournal of {F}luid {M}echanics\/} {\bf 395}, 211--236.

\bibitem[Srivatsan \& Kumaran(1997)]{srivatsan_kumaran_1997a}
{\sc Srivatsan, L \& Kumaran, V} 1997 Flow induced instability of the interface
  between a fluid and a gel. {\em {J}ournal de {P}hysique {II}\/} {\bf 7}~(6),
  947--963.

\bibitem[Suga {\em et~al.\/}(2010)Suga, Matsumura, Ashitaka, Tominaga \&
  Kaneda]{suga_matsumura_ashitaka_tominaga_kaneda_2010a}
{\sc Suga, K, Matsumura, Y, Ashitaka, Y, Tominaga, S \& Kaneda, M} 2010 Effects
  of wall permeability on turbulence. {\em {I}nternational {J}ournal of {H}eat
  and {F}luid {F}low\/} {\bf 31}, 974--984.

\bibitem[Sugiyama {\em et~al.\/}(2011)Sugiyama, Ii, Takeuchi, Takagi \&
  Matsumoto]{sugiyama_ii_takeuchi_takagi_matsumoto_2011a}
{\sc Sugiyama, K, Ii, S, Takeuchi, S, Takagi, S \& Matsumoto, Y} 2011 A full
  {E}ulerian finite difference approach for solving fluid--structure coupling
  problems. {\em {J}ournal of {C}omputational {P}hysics\/} {\bf 230}~(3),
  596--627.

\bibitem[Sussman {\em et~al.\/}(1994)Sussman, Smereka \&
  Osher]{sussman_smereka_osher_1994a}
{\sc Sussman, M, Smereka, P \& Osher, S} 1994 A level set approach for
  computing solutions to incompressible two-phase flow. {\em {J}ournal of
  {C}omputational {P}hysics\/} {\bf 114}~(1), 146--159.

\bibitem[Takeuchi {\em et~al.\/}(2010)Takeuchi, Yuki, Ueyama \&
  Kajishima]{takeuchi_yuki_ueyama_kajishima_2010a}
{\sc Takeuchi, S, Yuki, Y, Ueyama, A \& Kajishima, T} 2010 A conservative
  momentum-exchange algorithm for interaction problem between fluid and
  deformable particles. {\em {I}nternational {J}ournal for {N}umerical
  {M}ethods in {F}luids\/} {\bf 64}~(10-12), 1084--1101.

\bibitem[Tilton \& Cortelezzi(2006)]{tilton_cortelezzi_2006a}
{\sc Tilton, Nils \& Cortelezzi, Luca} 2006 The destabilizing effects of wall
  permeability in channel flows: A linear stability analysis. {\em {P}hysics of
  {F}luids (1994-present)\/} {\bf 18}~(5), 051702.

\bibitem[Tilton \& Cortelezzi(2008)]{tilton_cortelezzi_2008a}
{\sc Tilton, N \& Cortelezzi, L} 2008 Linear stability analysis of
  pressure-driven flows in channels with porous walls. {\em {J}ournal of
  {F}luid {M}echanics\/} {\bf 604}, 411--445.

\bibitem[Tryggvason {\em et~al.\/}(2007)Tryggvason, Sussman \&
  Hussaini]{tryggvason_sussman_hussaini_2007a}
{\sc Tryggvason, G, Sussman, M \& Hussaini, M~Y} 2007 Immersed boundary methods
  for fluid interfaces. {\em {C}omputational {M}ethods for {M}ultiphase
  {F}low\/} {\bf 3}.

\bibitem[Verma \& Kumaran(2013)]{verma_kumaran_2013a}
{\sc Verma, M K~S \& Kumaran, V} 2013 A multifold reduction in the transition
  reynolds number, and ultra-fast mixing, in a micro-channel due to a dynamical
  instability induced by a soft wall. {\em {J}ournal of {F}luid {M}echanics\/}
  {\bf 727}, 407--455.

\bibitem[Zalesak(1979)]{zalesak_1979a}
{\sc Zalesak, S~T} 1979 Fully multidimensional flux-corrected transport. {\em
  {J}ournal of {C}omputational {P}hysics\/} {\bf 31}, 335--362.

\end{thebibliography}

\end{document}